\documentclass[10pt,conference]{IEEEtran}

\usepackage[utf8]{inputenc}
\usepackage{amsmath}
\usepackage{amssymb,amsfonts}
\usepackage{floatrow}
\usepackage{graphicx}
\usepackage{textcomp}
\usepackage[dvipsnames]{xcolor}
\usepackage{booktabs}
\usepackage{algorithm}
\usepackage[noend]{algpseudocode}
\usepackage{listings}
\usepackage{inconsolata}
\usepackage{enumitem}
\usepackage{subcaption}
\usepackage{balance}
\usepackage{forest}
\usepackage{soul}
\usepackage{bbold}
\usepackage{hyperref}
\newcommand{\hlc}[2][yellow]{{%
    \colorlet{foo}{#1}%
    \sethlcolor{foo}\hl{#2}}%
}
\usepackage{pifont}
\newcommand{\cmark}{\ding{51}}%
\newcommand{\xmark}{\ding{55}}%

\newcommand{\boxalign}[2][0.9\columnwidth]{
  \par\noindent\tikzstyle{mybox} = [draw=black,inner sep=6pt]
  \begin{center}\begin{tikzpicture}
   \node [mybox] (box){%
    \begin{minipage}{#1}{\vspace{-5mm}#2}\end{minipage}
   };
  \end{tikzpicture}\end{center}
}

\definecolor{clcolor}{rgb}{0.5,0.7,0.9}
\definecolor{kscolor}{rgb}{0.9,0.1,0.1}
\definecolor{nkcolor}{rgb}{0.4,0.9,0.4}
\definecolor{rbcolor}{rgb}{0.9,0.4,0.96}

\definecolor{lightyellow}{HTML}{EEDD88}
\definecolor{mint}{HTML}{44BB99}
\definecolor{lightblue}{HTML}{77AADD}
\definecolor{orange}{HTML}{EE8866}

\newcommand\examples[0]{\ensuremath{S}}
\newcommand\oracle[0]{\ensuremath{\mathcal{O}}}
\newcommand\grammar[0]{\ensuremath{\mathcal{G}}}
\newcommand\trees[0]{\ensuremath{\mathcal{T}}}
\newcommand{\mathcolorbox}[2]{\colorbox{#1}{$\displaystyle #2$}}
\newcommand{\term}[1]{\textcolor{NavyBlue}{\texttt{#1}}}
\newcommand{\gramor}[0]{~|~}
\newcommand{\regexformat}[1]{\textcolor{BrickRed}{\footnotesize \textsf{#1}}}
\newcommand{\regexfmt}[1]{\textcolor{BrickRed}{\small \textsf{#1}}}
\newcommand{\ourmethod}[0]{\textsc{Arvada}}
\newcommand{\acceptbubble}[0]{\textsc{CheckBubble}}
\newcommand{\getbubbles}[0]{\textsc{GetBubbles}}
\newcommand{\replaces}[0]{\textsc{Replaces}}
\newcommand{\merges}[0]{\textsc{Merges}}

\title{Learning Highly Recursive Input Grammars}
 \author{\IEEEauthorblockN{Neil Kulkarni\textsuperscript{*}}
 \IEEEauthorblockA{University of California, Berkeley\\
 neil.kulkarni@berkeley.edu}
 \and
 \IEEEauthorblockN{Caroline Lemieux\textsuperscript{*}}
 \IEEEauthorblockA{University of California, Berkeley\\
 clemieux@cs.berkeley.edu}
  \and
 \IEEEauthorblockN{Koushik Sen}
 \IEEEauthorblockA{University of California, Berkeley\\
 ksen@cs.berkeley.edu}}

\begin{document}
\maketitle
\begingroup\renewcommand\thefootnote{*}
\footnotetext{Equal contribution.}
\endgroup

\begin{abstract}
This paper presents \ourmethod{}, an algorithm for learning context-free grammars from a set of positive examples and a Boolean-valued oracle. \ourmethod{} learns a context-free grammar by building parse trees from the positive examples. Starting from initially flat trees, \ourmethod{} builds structure to these trees with a key operation: it \emph{bubbles} sequences of sibling nodes in the trees into a new node, adding a layer of indirection to the tree. Bubbling operations enable recursive generalization in the learned grammar. We evaluate \ourmethod{} against GLADE and find it achieves on average increases of 4.98$\times$ in recall and 3.13$\times$ in F1 score, while incurring only a 1.27$\times$ slowdown and requiring only 0.87$\times$ as many calls to the oracle. \ourmethod{} has a particularly marked improvement over GLADE on grammars with highly recursive structure, like those of programming languages.  
\end{abstract}

\section{Introduction}
\label{sec:intro}

Learning a high-level language description from a set of examples in that language is a long-studied and difficult problem. While early interest in this problem was motivated by the desire to automatically learn human languages from examples, more recently the problem has been of interest in the context of learning program input languages. Learning a language of program inputs has several relevant applications, including generation of randomized test inputs~\cite{GopinathF1ArXiV2019,AschermannNautilusNDSS2019,Wang19}, as well as providing a high-level specification of inputs, which can aid both comprehension and debugging. 


In this paper we focus on the problem of learning \emph{context-free grammars} (CFGs) from a set of positive examples $S$ and a Boolean-value oracle \oracle{}. This is a similar setting as GLADE~\cite{BastaniGladePLDI2017}. Like GLADE, and unlike other recent related works~\cite{WuReinamFSE2019,HoscheleAutogramASE2016,GopoinathMimidFSE2020}, we assume the oracle is black-box: our technique can only see the Boolean return value of the oracle. 
We adopted the use of an oracle as we believe that in practice, an oracle---e.g. in the form of a parser---is easier to obtain than good, information-carrying negative examples.

In this paper, we describe a novel algorithm, \ourmethod{}, for learning CFGs from  example strings $S$ and an oracle \oracle{}. At a high-level, \ourmethod{} attempts to create the smallest CFG possible that accommodates all the examples. It uses two key operations---bubbling and merging---to generalize the language as much as possible, while not overgeneralizing beyond the language accepted by \oracle{}.

To create this context-free grammar, \ourmethod{} repeatedly performs the bubbling and merging operations on tree representations of the input examples. This set of trees is initialized with one ``flat'' tree per input example, i.e. the tree with a single root node whose children are the characters of the input string. The \emph{bubbling} operation takes sequences of sibling nodes in the trees and adds a layer of indirection by replacing the sequence with a new node. This new node has the bubbled sequence of sibling nodes as children.

Then \ourmethod{} decides whether to accept or reject the proposed bubble by checking whether a relabeling of the new node enables sound generalization of the learned language. Essentially, labels of non-leaf nodes correspond to nonterminals in the learned grammar.  Merging the labels of two distinct nodes in the trees adds new strings to the grammar's language: the strings derivable from  subtrees with the same label can be swapped. We call this the \emph{merge} operation since it merges the labels of two nodes in the tree. If a valid merge occurs, the structure introduced by the bubble is preserved.  Thus, merges introduce recursion when a parent node is merged with one of its descendants. If the label of the new node added in the bubbling operation cannot merge with any existing node in the trees, the bubble is rejected. That is, the introduced indirection node is removed, and the bubbled sequence of sibling nodes is restored to its original parent. These operations are repeated until no remaining bubbled sequence enables a valid merge.

In this paper, we formalize this algorithm in \ourmethod{}. We introduce heuristics in the ordering of bubble sequences minimize the number of bubbles \ourmethod{} must check before find a successful relabeling.
We implement \ourmethod{} in 2.2k LoC in Python, and make it available as open-source. We compare \ourmethod{} to GLADE~\cite{BastaniGladePLDI2017}, a state-of-the-art for grammar learning engine with blackbox oracles. We evaluate it on parsers for several grammars taken from the evaluation of GLADE, Reinam~\cite{WuReinamFSE2019}, Mimid~\cite{GopoinathMimidFSE2020}, as well as a few new highly-recursive grammars. On average across these benchmarks,  \ourmethod{} achieves $4.98\times$ higher recall and $3.13\times$ higher F1 score over GLADE. \ourmethod{} incurs on a slowdown of $1.27\times$ over GLADE, while requiring $0.87\times$ as many oracle calls. We believe this slowdown is reasonable, especially given the difference in implementation language---\ourmethod{} is implemented in Python, while GLADE is implemented in Java. Our contributions are as follows:
\begin{itemize}
\item We introduce \ourmethod{}, which learns grammars from inputs strings and oracle via bubble-and-merge operations.
\item We distribute \ourmethod{}'s implementation as open source: \url{https://github.com/neil-kulkarni/arvada}.
\item We evaluate \ourmethod{} on a variety of benchmarks against the state-of-the-art method GLADE. 
\end{itemize}

\section{Motivating Example}
\label{sec:motivation}

\begin{figure}
      \hspace{1em}  \fbox{$\grammar{}_w$} \hfill
    \vspace{-3em}
        \centering
    \boxalign{{ \small
    \begin{align*}
        \textit{start} \to&~ \textit{stmt} \\
        \textit{stmt}  \to&~ \term{while\textvisiblespace}~\textit{boolexpr}~\term{\textvisiblespace{}do\textvisiblespace{}}~ \textit{stmt} \\
        |&~ \term{if\textvisiblespace{}}~\textit{boolexpr}~\term{\textvisiblespace{}then\textvisiblespace{}}~ \textit{stmt} ~\term{\textvisiblespace{}else\textvisiblespace{}}~ \textit{stmt}\\
         |&~ \term{L\textvisiblespace{}=\textvisiblespace{}}~\textit{numexpr}\\
         |&~ \textit{stmt}~\term{\textvisiblespace{};\textvisiblespace{}}~\textit{stmt}\\
        \textit{boolexpr}  \to&~  \term{$\sim$}\textit{boolexpr}~|~\textit{boolexpr}~\term{\textvisiblespace{}\&\textvisiblespace{}}~\textit{boolexpr}\\
        |&~ \textit{numexpr}~\term{\textvisiblespace{}==\textvisiblespace{}}~\textit{numexpr}~|~\term{false}~|~\term{true}\\
        \textit{numexpr}  \to&~ \term{(}\,\textit{numexpr}\,\term{+}\,\textit{numexpr}\,\term{)}~|~\term{L}~|~\term{n}
    \end{align*}\vspace{-1.5em}}
    }
    \vspace{-1.5em}
    \hspace{-4.2em}
\begin{minipage}[t]{0.24\columnwidth}
\small
{
    \begin{align*}
	S=\{&\text{``}\texttt{while}\; \texttt{true}\; \texttt{\&}\; \texttt{false}\; \texttt{do}\; \texttt{L}\;\texttt{=}\;\texttt{n}\text{''}, \\
    &\text{``}\texttt{L = n ; L = (n+n)}\text{''}\}
    \end{align*}
}
\end{minipage}
\begin{minipage}[t]{0.24\columnwidth}
\small
    $$ \oracle(i) = \begin{cases}
   \texttt{True}\; \text{if}\; i\in \mathcal{L}(\grammar_w)\\
   \texttt{False}\; \text{otherwise}
    \end{cases}$$
\end{minipage}
\vspace{-1.7em}

    \caption{Example inputs $S$, and oracle $\mathcal{O}$ which returns \texttt{true} if its input is in the language of  the while grammar $\mathcal{G}_w$. }
    \label{fig:motivating-example}
\end{figure}

\ourmethod{} takes as input a set of example strings $S$ and an oracle \oracle{}. The oracle returns \texttt{True} if its input string is valid and \texttt{False} otherwise. \ourmethod{}'s goal is to learn a grammar \grammar{} which maximally generalizes the example strings $S$ in a manner \emph{consistent} with the oracle \oracle{}. That is, strings $i\in\mathcal{L}(\grammar{})$ in the language of the learned grammar should with high probability be accepted by the oracle: $\oracle(i) = \texttt{True}$. We formally describe maximal generalization in  Section~\ref{sec:technique}.

Fundamentally, \ourmethod{} learns a grammar by learning ``parse trees'' for the examples in $S$. These parse trees are initialized with flat trees for each example in $S$. Then, \ourmethod{} adds structure, turning sequences of sibling nodes into new subtrees. The particular subtrees \ourmethod{} keeps are those which enable generalization in the induced grammar.

From any set of trees \trees{} we can derive an \textbf{\emph{induced grammar}}. In particular, each non-leaf node in a tree $t\in\trees{}$ with label $t_\textit{parent}$ and children with labels $t_{\textit{child}_1}, t_{\textit{child}_2}, \dots, t_{\textit{child}_n}$ \textbf{\emph{induces the rule}} $t_\textit{parent}\to t_{\textit{child}_1} t_{\textit{child}_2} \cdots t_{\textit{child}_n}$. The \emph{induced grammar} of \trees{} is then the set of induced rules for all nodes in the trees. 

For example, the trees in Fig.~\ref{fig:motivate-start} induce the grammar: 
\vspace{-0.2em}
{\small
\begin{align*}
&t_0 \to \texttt{w\;h\;i\;l\;e\;\textvisiblespace{}\;t\;r\;u\;e\;\textvisiblespace{}\;\&\;f\;a\;l\;s\;e\;\textvisiblespace{}\;d\;o\;\textvisiblespace{}\;L\;\textvisiblespace{}\;=\;\textvisiblespace{}\;n}\\
&t_0 \to \texttt{L\;\textvisiblespace{}\;=\;\textvisiblespace{}\;n\;\textvisiblespace{}\;;\;\textvisiblespace{}\;L\;\textvisiblespace{}\;=\;\textvisiblespace{}\;(\;n\;+\;n\;)}
\end{align*}
}
\vspace{-0.2em}
and the trees under (4) in Fig.~\ref{fig:motivate-full-run} induce the grammar in Fig.~\ref{fig:motivate-grammar}.

Because of this mapping from trees to grammars, we will use the term ``nonterminal'' interchangeably with ``label of a non-leaf node'' when discussing relabeling trees. 

\subsection{Walkthrough}
We illustrate \ourmethod{} on a concrete example. We take the set of examples $S$ and oracle \oracle{}  shown in Fig.~\ref{fig:motivating-example}. This oracle \oracle{} accepts inputs as valid only if they are in the language of the while grammar $\grammar_w$, shown at the top of the figure. \ourmethod{} treats $\mathcal{O}$ as blackbox, that is, it has no structural knowledge of $\grammar_w$: $\grammar_w$ is shown only to clarify the behavior of \oracle{}.

\begin{figure}
	\centering
	\resizebox{\columnwidth}{!}{
	\begin{forest} [$t_0$, l sep={1.2cm}, for tree= {s sep={0.25mm}} [\texttt{w}] [\texttt{h}] [\texttt{i}] [\texttt{l}] [\texttt{e}] [\textvisiblespace{}] [\texttt{t}] [\texttt{r}] [\texttt{u}] [\texttt{e}] [\textvisiblespace{}] [\texttt{\&}] [\texttt{f}] [\texttt{a}] [\texttt{l}] [\texttt{s}] [\texttt{e}] [\textvisiblespace{}] [\texttt{d}] [\texttt{o}] [\textvisiblespace{}] [\texttt{L}] [\textvisiblespace{}] [\texttt{=}] [\textvisiblespace{}] [\texttt{n}]]	\end{forest} }
\scalebox{0.8}{
	\begin{forest}
		[$t_0$, l sep={0.6cm}, for tree= {s sep={0.5mm}} [\texttt{L}] [\textvisiblespace{}] [\texttt{=}] [\textvisiblespace{}] [\texttt{n}] [\textvisiblespace{}] [\texttt{;}]  [\textvisiblespace{}] [\texttt{L}] [\textvisiblespace{}][\texttt{=}] [\textvisiblespace{}] [\texttt{(}] [\texttt{n}] [\texttt{+}] [\texttt{n}] [\texttt{)}]]
	\end{forest}
}
 	\caption{Initial set of parse trees \trees{} created by \ourmethod{} when run on $S$, \oracle{} in Fig.~\ref{fig:motivating-example}. Each terminal $c$ has a nonterminal parent $t_c$ with rule $t_c \rightarrow c$, omitted for simplicity.}
	\label{fig:motivate-start}
\end{figure}
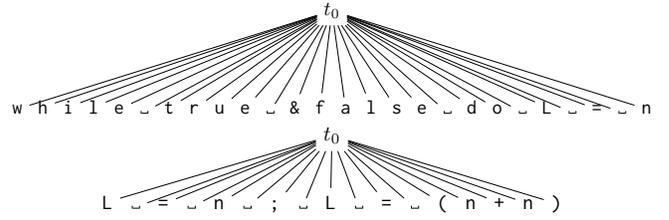

\ourmethod{} begins by constructing na\"ive, flat, parse trees from the examples. These are shown in Fig.~\ref{fig:motivate-start}. Essentially, these trees simply go from the start nonterminal $t_0$ to the sequence of characters in each example $s\in S$. Let \trees{} designate the set of trees \ourmethod{} maintains at any point in its algorithm. 

\subsubsection{Bubbling}
\begin{figure}
\centering
Bubble $t_1\to  {\small \texttt{h i l e}}$ {\Large \xmark{}}
\resizebox{0.9\columnwidth}{!}{
	\begin{forest} baseline [$t_0$, l sep={0.75cm}, for tree= {s sep={0.25mm}} [\texttt{w}] 
		[$t_1$, for tree={fill=mint} [\texttt{h}] [\texttt{i}] [\texttt{l}] [\texttt{e}]] [\textvisiblespace{}] [\texttt{t}] [\texttt{r}] [\texttt{u}] [\texttt{e}] [\textvisiblespace{}] [\texttt{\&}] [\texttt{f}] [\texttt{a}] [\texttt{l}] [\texttt{s}] [\texttt{e}] [\textvisiblespace{}] [\texttt{d}] [\texttt{o}] [\textvisiblespace{}] [\texttt{L}] [\textvisiblespace{}] [\texttt{=}] [\textvisiblespace{}] [\texttt{n}]]
\end{forest}
}

\vspace{0.6em}
\rule{0.8\columnwidth}{0.4pt}
\vspace{0.6em}

Bubble $t_2\to  {\small \texttt{( n + n )}}$ {\Large \cmark{}}
\scalebox{0.8}{
\begin{forest}
[$t_0$, l sep={0.6cm}, for tree= {s sep={0.5mm}} [\texttt{L}] [\textvisiblespace{}] [\texttt{=}] [\textvisiblespace{}] [\texttt{n}] [\textvisiblespace{}] [\texttt{;}]  [\textvisiblespace{}] [\texttt{L}] [\textvisiblespace{}][\texttt{=}] [\textvisiblespace{}] [$t_2$, l sep={0.3cm},for tree={fill=mint}, [\texttt{(}] [\texttt{n}] [\texttt{+}] [\texttt{n}] [\texttt{)}]]]
\end{forest}
}

\caption{Two possible bubbles applied to the trees in Fig.~\ref{fig:motivate-start}.}
\label{fig:motivate-single-bubbles}
\end{figure}
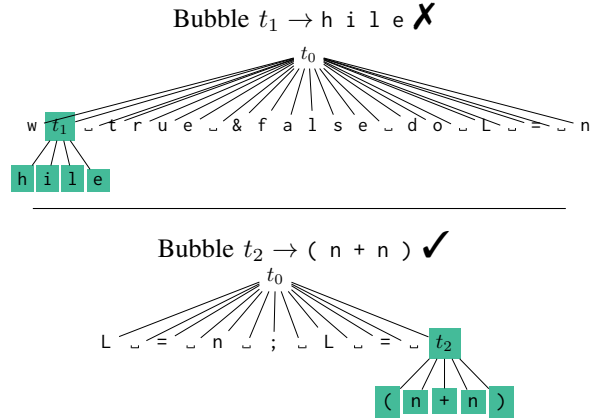

The fundamental operation \ourmethod{} performs is to \emph{bubble up} a sequence of sibling nodes in the current trees \trees{} into a new nonterminal.  To bubble a sequence $s_1$ in the trees \trees{}, we create a new nonterminal node $t_{s_1}$ with children $s_1$. Then we replace all occurrences of $s_1$ in each $t\in\trees{}$ with $t_{s_1}$. Fig.~\ref{fig:motivate-single-bubbles} shows two such bubbles applied to the trees in Fig.~\ref{fig:motivate-start}. On top, we have bubbled the sequence \texttt{hile} into $t_1$; the second tree, unchanged, is not illustrated. On the bottom, we have bubbled \texttt{(n+n)} into $t_2$; the first tree is unchanged.

\subsubsection{Merging}
After bubbling a sequence $s_1$, \ourmethod{} either \emph{accepts} or \emph{rejects} the bubble. \ourmethod{} only accepts a bubble if it enables valid generalization of the examples. That is, if a relabeling of the bubbled nonterminal---merging its label with the label of another existing node---expands the language accepted by the induced grammar, while maintaining the oracle-validity of the strings produced by the induced grammar.

Consider again Fig.~\ref{fig:motivate-single-bubbles}. On top, we have the bubble $t_1\to  \texttt{hile}$. There is no terminal or nonterminal whose label can be merged with the label $t_1$ and retain a valid grammar: it can't be merged with $t_0$, since ``\texttt{hile}'' on its own is not accepted by \oracle. Nor can it be merged with the label of any individual character: as just one example, merging with \texttt{L} would cause the \oracle{}-invalid generalization ``\texttt{hile = n ; hile = (n+n)}''.

\begin{figure}
	\centering
	\resizebox{\columnwidth}{!}{
	\begin{forest} [$t_0$, l sep={1.2cm}, for tree= {s sep={0.25mm}} [\texttt{w}] [\texttt{h}] [\texttt{i}] [\texttt{l}] [\texttt{e}] [\textvisiblespace{}] [\texttt{t}] [\texttt{r}] [\texttt{u}] [\texttt{e}] [\textvisiblespace{}] [\texttt{\&}] [\texttt{f}] [\texttt{a}] [\texttt{l}] [\texttt{s}] [\texttt{e}] [\textvisiblespace{}] [\texttt{d}] [\texttt{o}] [\textvisiblespace{}] [\texttt{L}] [\textvisiblespace{}] [\texttt{=}] [\textvisiblespace{}] [\texttt{n}]]	\end{forest} 
\begin{forest}
	[$t_0$, l sep={1cm}, for tree= {s sep={0.5mm}} [\texttt{L}] [\textvisiblespace{}] [\texttt{=}] [\textvisiblespace{}] [\texttt{n}] [\textvisiblespace{}] [\texttt{;}]  [\textvisiblespace{}] [\texttt{L}] [\textvisiblespace{}][\texttt{=}] [\textvisiblespace{}] [\texttt{(}] [\texttt{n}] [\texttt{+}] [\texttt{n}] [\texttt{)}]]
\end{forest}}
 	\vspace{-.3em}
 $$|$$ 
 
 \vspace{-1em}
{ \small(1) Bubble $t_2\to  \texttt{(n+n)}$; merge  ($t_2$, \texttt{n}) into $t_3$}
 \vspace{-.5em}
 $$\downarrow$$
 \vspace{-2em}
 	\resizebox{\columnwidth}{!}{
 		\begin{forest} 
 			[$t_0$, l sep={1cm}, for tree= {s sep={0.25mm}} 
 			[\texttt{w}] 
 			[\texttt{h}] 
 			[\texttt{i}] 
 			[\texttt{l}] 
 			[\texttt{e}] 
 			[\textvisiblespace{}] 
 			[\texttt{t}] 
 			[\texttt{r}] 
 			[\texttt{u}] 
 			[\texttt{e}]
 			[\textvisiblespace{}]  
 			[\texttt{\&}] 
 			[\textvisiblespace{}] 
 			[\texttt{f}] 
 			[\texttt{a}]
 			 [\texttt{l}] 
 			 [\texttt{s}]
 			  [\texttt{e}]
 			  [\textvisiblespace{}] 
 			  [\texttt{d}] 
 			  [\texttt{o}] 
 			  [\textvisiblespace{}] 
 			  [\texttt{L}]
 			  [\textvisiblespace{}]  
 			  [\texttt{=}] 
 			  [\textvisiblespace{}] 
 			  [$t_3$,  fill=mint [\texttt{n}, l =0.75mm]]
 			]
 		\end{forest}
 		\begin{forest}
 			[$t_0$, l sep ={0.75cm}, for tree= {s sep={0.5mm},  l sep = {3mm}, l =1mm} 
 			[\texttt{L}] 
 			[\textvisiblespace{}] 
 			[\texttt{=}] 
 			[\textvisiblespace{}] 
 			[$t_3$, fill=mint [\texttt{n} ]
 			] 
 			[\textvisiblespace{}] 
 			[\texttt{;}] 
 			[\textvisiblespace{}] 
 			[\texttt{L}] 
 			[\textvisiblespace{}] 
 			[\texttt{=}] 
 			[\textvisiblespace{}] 
 			[$t_3$, fill=mint, 
 				[\texttt{(}] 
 				[$t_3$, fill=mint
 				[\texttt{n}]] 
 				[\texttt{+}] 
 				[$t_3$, fill=mint 
 					[\texttt{n}]
 					] 
 				[\texttt{)}]
 				]
 			]
 		\end{forest}
 	}
 	\vspace{-1em}
 $$|$$ 
 
 \vspace{-1em}
 {\small (2) Bubble $t_4 \to \texttt{L\textvisiblespace=\textvisiblespace}~t_3$; merge  ($t_4$, $t_0$) into $t_0$}
 \vspace{-.5em}
 $$\downarrow$$
 \vspace{-2em}
 \resizebox{\columnwidth}{!}{
 	\begin{forest} 
 		[$t_0$, fill=mint,l sep={1cm}, for tree= {s sep={0.25mm}} 
 			[\texttt{w}] 
 			[\texttt{h}] 
 			[\texttt{i}] 
 			[\texttt{l}] 
 			[\texttt{e}] 
 			[\textvisiblespace{}] 
 			[\texttt{t}] 
 			[\texttt{r}] 
 			[\texttt{u}] 
 			[\texttt{e}] 
 			[\textvisiblespace{}] 
 			[\texttt{\&}] 
 			[\textvisiblespace{}] 
 			[\texttt{f}] 
 			[\texttt{a}]
 			[\texttt{l}] 
 			[\texttt{s}]
 			[\texttt{e}] 
 			[\textvisiblespace{}] 
 			[\texttt{d}] 
 			[\texttt{o}]  
 			[\textvisiblespace{}] 
 			[$t_0$, fill = mint
 				[\texttt{L}] 
 				[\textvisiblespace{}] 
 				[\texttt{=}] 
 				[\textvisiblespace{}] 
 				[$t_3$,[\texttt{n}, l=0.75mm]]
 			]
 		]
 	\end{forest}
 	\begin{forest}
 	[$t_0$,fill=mint, for tree= {s sep={0.5mm},  l sep = {2mm}, l =1mm} 
 	[$t_0$ 		, fill = mint
 	[\texttt{L}] 
 	[\textvisiblespace{}] 
 	[\texttt{=}] 
 	[\textvisiblespace{}] 
 	[$t_3$, 
 	[\texttt{n}, l=0.75mm]] 
 	]
 	[\textvisiblespace{}] 
 	[\texttt{;}] 
 	[\textvisiblespace{}] 
 	[$t_0 $		, fill = mint		
 	[\texttt{L}] 
 	[\textvisiblespace{}] 
 	[\texttt{=}] 
 	[\textvisiblespace{}] 
 	[$t_3$, 
 	[\texttt{(}] 
 	[$t_3$, [\texttt{n}, l=0.75mm]] 
 	[\texttt{+}] 
 	[$t_3$,  [\texttt{n}, l=0.75mm]] 
 	[\texttt{)}]
 	]
 	]
 	]
 \end{forest}
 }
 	\vspace{-2em}
 $$|$$ 
 
 \vspace{-1em}
{\small 
	(3) 2-Bubble ($t_5 \to \texttt{false}$, $t_6 \to \texttt{true}$); merge both into $t_7$}
 \vspace{-.5em}
 $$\downarrow$$
 
 \vspace{-1em}
 \scalebox{0.6}{
 	\begin{forest} 
 		[$t_0$,  for tree= {s sep={0.25mm}} 
 		[\texttt{w}] 
 		[\texttt{h}] 
 		[\texttt{i}] 
 		[\texttt{l}] 
 		[\texttt{e}]
 		[\textvisiblespace]
 		[$t_7$, fill=mint
 			[\texttt{t}] 
 			[\texttt{r}] 
 			[\texttt{u}] 
 			[\texttt{e}] 
 		]
 		[\textvisiblespace]
 		[\texttt{\&}] 
 		[\textvisiblespace]
 		[$t_7$, fill=mint
 			[\texttt{f}] 
 			[\texttt{a}]
 			[\texttt{l}] 
 			[\texttt{s}]
 			[\texttt{e}] 
 		]
 		[\textvisiblespace]
 		[\texttt{d}] 
 		[\texttt{o}]  
 		[\textvisiblespace]
 		[$t_0$
 		[\texttt{L}] 
 		[\textvisiblespace]
 		[\texttt{=}] 
 		[\textvisiblespace]
 		[$t_3$,[\texttt{n}]]
 		]
 		]
 	\end{forest}
 		\begin{forest}
	[$t_0$, for tree= {s sep={0.5mm},  l sep = {2mm}, l =1mm} 
	[$t_0$ 		
	[\texttt{L}] 
	[\textvisiblespace]
	[\texttt{=}] 
	[\textvisiblespace]
	[$t_3$, 
	[\texttt{n}]] 
	]
	[\textvisiblespace]
	[\texttt{;}] 
	[\textvisiblespace]
	[$t_0 $		
	[\texttt{L}] 
	[\textvisiblespace]
	[\texttt{=}] 
	[\textvisiblespace]
	[$t_3$, 
	[\texttt{(}] 
	[$t_3$, [\texttt{n}]] 
	[\texttt{+}] 
	[$t_3$,  [\texttt{n}]] 
	[\texttt{)}]
	]
	]
	]
\end{forest}
 }	
\vspace{-3em}

 	$$|$$ 
 
 	\vspace{-1em}
 {\small (4) Bubble $t_8 \to t_7\texttt{ \textvisiblespace{} \& \textvisiblespace{} } t_7$; merge  ($t_8$, $t_7$) into $t_9$}
 	\vspace{-.5em}
 	$$\downarrow$$
 	
 	\vspace{-1em}
 \scalebox{0.7}{
 		\begin{forest} 
 			[$t_0$,  for tree= {s sep={0.25mm}} 
 			[\texttt{w}] 
 			[\texttt{h}] 
 			[\texttt{i}] 
 			[\texttt{l}] 
 			[\texttt{e}]
 			[\textvisiblespace]
 			[$t_9$, fill=mint
 				[$t_9$, fill=mint
 					[\texttt{t}] 
 					[\texttt{r}] 
 					[\texttt{u}] 
 					[\texttt{e}] 
 				]
 				 			[\textvisiblespace]
 				[\texttt{\&}] 
 				 			[\textvisiblespace]
 				[$t_9$, fill=mint
 					[\texttt{f}] 
 					[\texttt{a}]
 					[\texttt{l}] 
 					[\texttt{s}]
 					[\texttt{e}] 
 				]
 			]
 			[\textvisiblespace]
 			[\texttt{d}] 
 			[\texttt{o}]  
 			[\textvisiblespace]
 			[$t_0$
 			[\texttt{L}] 
 			 			[\textvisiblespace]
 			[\texttt{=}] 
 			 			[\textvisiblespace]
 			[$t_3$,[\texttt{n}]]
 			]
 			]
 		\end{forest}
 		\begin{forest}
 			[$t_0$, for tree= {s sep={0.5mm},  l sep = {2mm}, l =1mm} 
 			[$t_0$ 		
 			[\texttt{L}] 
 			[\textvisiblespace]
 			[\texttt{=}] 
 			[\textvisiblespace]
 			[$t_3$, 
 			[\texttt{n}]] 
 			]
 			[\textvisiblespace]
 			[\texttt{;}] 
 			[\textvisiblespace]
 			[$t_0 $		
 			[\texttt{L}] 
 			[\textvisiblespace]
 			[\texttt{=}] 
 			[\textvisiblespace]
 			[$t_3$, 
 			[\texttt{(}] 
 			[$t_3$, [\texttt{n}]] 
 			[\texttt{+}] 
 			[$t_3$,  [\texttt{n}]] 
 			[\texttt{)}]
 			]
 			]
 			]
 		\end{forest}
 }
	\caption{The state of trees \trees{} and the accepted bubbles of a full run of \ourmethod{} on $S$, \oracle{} in Fig.~\ref{fig:motivating-example}.}
	\label{fig:motivate-full-run}
\end{figure}
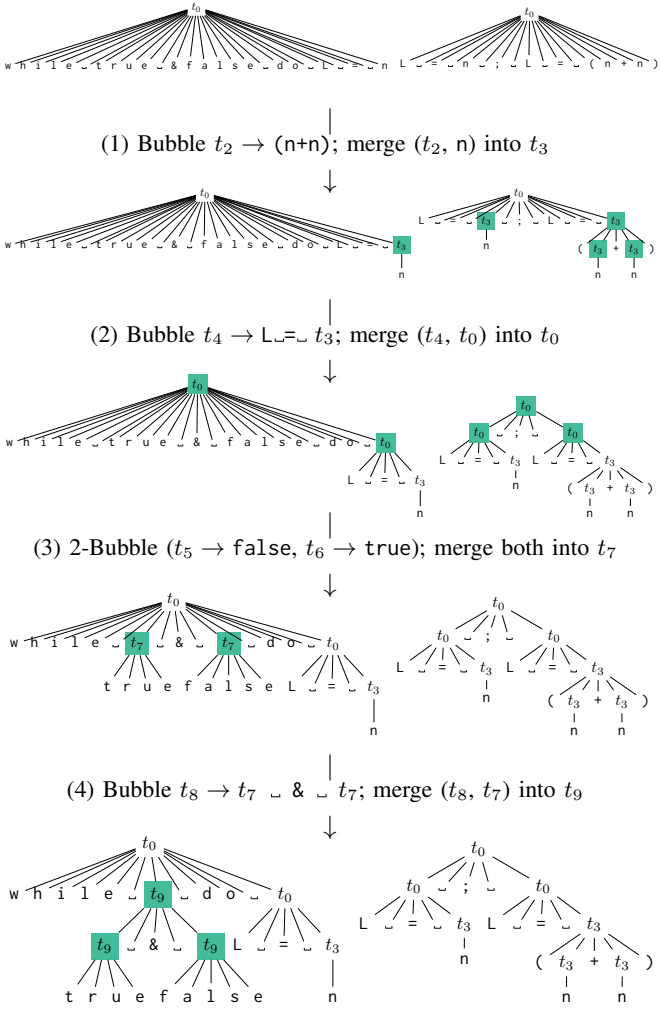

On the bottom of Fig~\ref{fig:motivate-single-bubbles}, we have the bubble $t_2\to  \texttt{(n+n)}$. We can in fact merge the label $t_2$ with the label $t_n$, the implicit nonterminal expanding to $\texttt{n}$. Notice that if we replace $\texttt{n}$ with the strings derivable from $t_2$, we get examples like \texttt{while true \& false do L = (n+n)} and \texttt{L = (n+n) ; L = ((n+n)+(n+n))}, which are all valid. Conversely, if we replace occurrences of $t_2$ with $\texttt{n}$, we get examples like \texttt{L = n ; L = n}. We accept this bubble, which expands the language accepted by the induced grammar. Thus, $t_2$ and \texttt{n} are merged and relabeled as $t_3$. The trees after the relabel are shown after (1) in Fig.~\ref{fig:motivate-full-run}. Note this merge has introduced recursive generalization; the induced grammar now includes the rules: 
\begin{align*}
	\small
t_0\to \texttt{L\textvisiblespace{}=\textvisiblespace}\, t_3 \quad t_0\to \texttt{L\textvisiblespace{}=\textvisiblespace{}}\, t_3 \,\texttt{\textvisiblespace{};\textvisiblespace{}}\, \texttt{L\textvisiblespace{}=\textvisiblespace{}}\, t_3 \quad t_3\to \texttt{(}t_3\texttt{+}t_3\texttt{)}\quad t_3\to \texttt{n}
\end{align*}

In practice, \ourmethod{} checks whether labels $t_a, t_b$ can be merged by checking \emph{candidate strings} against the oracle. If the oracle accepts all these candidate strings, the relabeling is valid and the labels are merged.  To create these candidates, \ourmethod{} creates mutated trees from the trees in \trees{} where  (1) subtrees rooted at  $t_a$ are replaced subtrees rooted at $t_b$, and (2) subtrees rooted at  $t_b$ are replaced subtrees rooted at $t_a$. The candidate strings are then the ones derived from these trees, i.e. the ordered sequence of a tree's leaf nodes. Section~\ref{sec:bubble-accepting}  describes the conditions under which a bubble is accepted in more detail. Section~\ref{sec:string-sampling} describes how to create these candidate strings, and the soundness issues this introduces.

\subsubsection{Double bubbling}
After accepting a bubble, \ourmethod{} continues to try and create new bubbles. It bubbles different sequences 
of children in the current trees \trees{}, checking if they are accepted, and updating \trees{} accordingly. Fig.~\ref{fig:motivate-full-run} shows a potential run of \ourmethod{}, with the state of the trees \trees{} as they are updated by bubbles and label merging.

In Fig.~\ref{fig:motivate-full-run}, after (1) accepting the bubble $t_2\to  \texttt{(n+n)}$, \ourmethod{} (2) finds and accepts the bubble  $t_4\to \texttt{L\textvisiblespace{}=\textvisiblespace{}}\; t_3$, whose label can be merged with the start nonterminal $t_0$. At this point,  \ourmethod{} will find no more bubbles which can be merged with any existing nodes in \trees{}. For example, if \ourmethod{} creates the bubble $t_5\to\texttt{true}$, it will find that the label $t_5$ cannot be merged with the label of any existing node and reject it.

To cope with this, \ourmethod{} also considers 2-bubbles. In a 2-bubble, two distinct sequences of children---say, $s_1$ and $s_2$---in the trees are bubbled at the same time, i.e. replacing both $s_1$ with $t_{s_1}\to s_1$ and some other $s_2$ with $t_{s_2}\to s_2$. The two sequences can be totally distinct, or sub/super sets, but not overlapping: ($s_1=\texttt{true}$, $s_2=\texttt{false}$) is ok, as is  ($s_1=\texttt{true}$, $s_2=\texttt{true \& false}$), but  ($s_1=$\texttt{ru\hlc[lightyellow]{e\textvisiblespace{}\&\textvisiblespace{}f}}, $s_2=$\texttt{\hlc[lightyellow]{e\textvisiblespace\&\textvisiblespace{}f}al}) is not. \ourmethod{} accepts a 2-bubble only if the labels $t_{s_1}$ and $t_{s_2}$ can be merged \emph{with each other}, not with another existing node. Otherwise, either  $t_{s_1}$ or $t_{s_2}$ could be accepted as a 1-bubble. 

\subsubsection{Termination}
In the run in Fig.~\ref{fig:motivate-full-run}, (3) \ourmethod{} applies and accepts the 2-bubble ($s_1=\texttt{true}$, $s_2=\texttt{false}$) and merges these sequences into $t_7$. This 2-bubble enables one final single bubble to be applied and accepted: (4) $t_8\to t_7\,\texttt{\textvisiblespace{}\&\textvisiblespace{}}\,t_7$ can be merged with $t_7$. After this, no more 1-bubbles or 2-bubbles can be accepted, 
so \ourmethod{} simply outputs the grammar induced by the final set of trees \trees{}.  Fig.~\ref{fig:motivate-grammar} shows the grammar.

\begin{figure}
\vspace{-1em}
	\centering
	\begin{align*}
	\textit{$t_0$} \to&~ \term{while\textvisiblespace{}}\ \textit{$t_9$}\ \term{\textvisiblespace{}do\textvisiblespace{}}\ \textit{$t_0$}~
	|~ \term{L\textvisiblespace{}=\textvisiblespace{}}\ \textit{$t_3$}~
	|~ \textit{$t_0$}\ \term{\textvisiblespace{};\textvisiblespace{}}\ \textit{$t_0$}\\
	\textit{$t_9$} \to&~ \textit{$t_9$}\ \term{\textvisiblespace{}\&\textvisiblespace{}}\ \textit{$t_9$}
	~|~ \term{true}
	~|~ \term{false}\\
	\textit{$t_3$} \to&~ \term{(}\,\textit{$t_3$}\,\term{+}\, \textit{$t_3$} \term{)}~|~ \term{n}
	\end{align*}
\vspace{-1em}
	\caption{Grammar produced by the run of \ourmethod{} in Fig.~\ref{fig:motivate-full-run}.}
	\label{fig:motivate-grammar}
	\vspace{-1em}
\end{figure}

\subsubsection{Effect of bubbling order}
First, note that multiple  orderings of bubbles can result in an equivalent grammar. For example, we could have applied ($s_1=\texttt{true}$, $s_2=\texttt{true\textvisiblespace{}\&\textvisiblespace{}false}$) in (3), then bubbled up \texttt{false} alone in (4). Second, while Fig.~\ref{fig:motivate-full-run} shows an ideal run, some accepted bubbles may impede further generalization of the grammar. For example, in the initial flat parse trees, $t_1\to\texttt{e\textvisiblespace{}\&\textvisiblespace{}false}$ can be merged with \texttt{e}. In the presence of the additional example ``\texttt{while n == n do skip}'', this merge prevents maximal generalization. 

As such, the order in which bubbles are applied and checked has a large impact on \ourmethod{}'s performance. In Section~\ref{sec:bubble-order}, we describe heuristics that order the bubbles for exploration based on the context and frequency of the bubbled subsequence. These heuristics increase \ourmethod{}'s probability of successfully finding the maximal generalization of $S$ with respect to \oracle{}, as discussed in Section~\ref{sec:eval-accuracy}.	

\subsubsection{Maximality of learned grammar}
The grammar in Fig.~\ref{fig:motivate-grammar} is not identical to that in Fig.~\ref{fig:motivating-example}. However, it contains all the rules in $\grammar_w$ demonstrated by the examples $S$:  $t_3$ has taken on the role of $\textit{numexpr}$, $t_9$ in the role of $\textit{boolexpr}$, and $t_0$ is effectively \textit{stmt}. However, the rule $ \textit{boolexpr} \to  \textit{numexpr}~\texttt{\textvisiblespace{}==\textvisiblespace{}}~\textit{numexpr}$ does not appear in Fig.~\ref{fig:motivate-grammar}. Fundamentally, this is because no substring derivable from this rule exists in $S$; as such, it is not part of $S$'s maximal generalization.

\section{Technique} 
\label{sec:technique}

We formally describe the high-level \ourmethod{} algorithm in Section~\ref{sec:main-algo}; Sections~\ref{sec:bubble-order},~\ref{sec:bubble-accepting}, ~\ref{sec:string-sampling}, and~\ref{sec:pre-tokenization} delve into the heuristic decisions made in \ourmethod{}'s implementation. 

First, we formalize our problem statement. \ourmethod{} accepts as input a set of example strings $S$ and a Boolean-valued oracle \oracle{} which judges the validity of the strings. \ourmethod{}'s goal is to learn a context-free grammar \grammar{} which \emph{maximally generalizes} the set of example  $S$ in a manner \emph{consistent} with \oracle{}. 


\paragraph*{Maximal generalization} Let $S$ be a set of input strings and \oracle{} be a Boolean-valued oracle accepting strings as input. Assume each $s\in S$ is accepted by the oracle, i.e., $ \forall s\in S\colon \oracle(s)=\texttt{True}$. Let $\grammar_\oracle$ be a context-free grammar such that its language of strings $\mathcal{L}(\grammar_\oracle)$ is equal to $\{i\in \Sigma^*~|~\oracle(i)=\texttt{True}\}$, the set of strings accepted by the oracle \oracle{}. Since $\oracle(s)=\texttt{True}$ for each $s\in S$, then each $s\in \mathcal{L}(\grammar_\oracle)$. We call $\mathcal{G_O}$ as the \emph{target grammar}.


Thus, for each $s$, there exists a derivation $\mathcal D_s$ from the start symbol $T_0$ to $s$, i.e. $\mathcal{D}_s = T_0\to \alpha_1\alpha_2\cdots\alpha_n \to \cdots \to s$. This derivation is a sequence of nonterminal expansions according to some rules $\grammar_\oracle$. Let $R_s$ be the set of rules in  $\grammar_\oracle$ used in the derivation $\mathcal{D}_s$. Let $R_S=\cup_{s\in S}R_s$, and $\grammar_\oracle^S$ be the subset of $\grammar_\oracle$ which contains only those rules $r\in R_S$. Intuitively, $\grammar_\oracle^S$ is the sub-grammar of $\grammar_\oracle$ which is exercised by the $s\in S$.

Finally: a grammar which \textbf{\emph{maximally generalizes} $S$ \emph{w.r.t.} \oracle{}} is a grammar \grammar{} such that $\mathcal{L}(\grammar{})= \mathcal{L}(\grammar_\oracle^S)$, i.e. it accepts the same language as $\grammar_\oracle^S$.

\subsection{Main Algorithm}
\label{sec:main-algo}

Algorithm~\ref{alg:full-grammar-learning} shows the main \ourmethod{} algorithm. It works as follows.
First, \ourmethod{} builds na\"ive, flat, parse trees from the input strings (Line~\ref{line:naive-trees}).
Considering each $s_i\in S$ as a sequence of characters  $s_i=c_i^1c_i^2\cdots c_i^{n_i}$, the tree constructed for $s_i$ has a root node with the start symbol label $t_0$ and $n_i$ children with labels $t_{c_i^1}, t_{c_i^2}, \dots, t_{c_i^{n_i}}$. Each $t_{c}$  has a single child whose label is the corresponding character $c$. Fig.~\ref{fig:motivate-start} shows these flat parse trees for the examples strings $s\in S$ in Fig.~\ref{fig:motivating-example}, although the $t_{c}\to c$ are not illustrated for simplicity.

\renewcommand{\algorithmicrequire}{\textbf{Input:}}
\renewcommand{\algorithmicensure}{\textbf{Output:}}
\begin{algorithm}
  \caption{ \ourmethod{}'s high-level algorithm}\label{alg:full-grammar-learning}
  \begin{algorithmic}[1]
  \algrenewcommand\algorithmicindent{0.7em}%
\Require a set of examples \examples{}, an language oracle \oracle{}.
\Ensure a grammar \grammar{} fitting the language.
    \State $\textit{bestTrees} \gets $ \Call{NaiveParseTrees}{\examples{}} \label{line:naive-trees}
    \State $\textit{bestTrees} \gets $ \Call{MergeAllValid}{\textit{bestTrees}, \oracle{}} \label{line:merge-all-start}
    \State $\textit{updated} \gets \texttt{True}$
    \While{\textit{updated}}
        \State $\textit{updated} \gets \texttt{False}$
        \State $\textit{allBubbles} \gets $ \Call{GetBubbles}{\textit{bestTrees}} \label{line:get-bubbles}
        \For{\textit{bubble} \textbf{in} \textit{allBubbles}} \label{line:bubble-loop}
            \State $\textit{bbldTrees} \gets $ \Call{Apply}{\textit{bestTrees}, \textit{bubble}} \label{line:apply-bubbles}
            \State $\textit{accepted}, \textit{mergedTs} \gets \Call{\acceptbubble}{\textit{bbldTrees}, \oracle}$\label{line:check-bubbles}
            \If{\textit{accepted}}
                \State $\textit{bestTrees} \gets \textit{mergedTs}$ \label{line:give-merged-tree}
                \State $\textit{updated} \gets \texttt{True}$
                \State \textbf{break}
            \EndIf
        \EndFor
    \EndWhile
    \State $\grammar \gets $ \Call{InducedGrammar}{\textit{bestTrees}}
    \State \Return \grammar{} \label{line:ret-main}
  \end{algorithmic}
\end{algorithm}

\ourmethod{} tries to generalize these parse trees by merging nodes in the tree into new nonterminal labels (Line~\ref{line:merge-all-start}). To merge two nodes $t_a$, $t_b$ in a tree, we replace all occurrences of the labels $t_a$, $t_b$ with a new label $t_c$. This creates new trees $\trees{}'$; the merge is valid if the language of the induced grammar of $\trees{}'$ only includes strings accepted by the oracle \oracle{}.

In practice, we 
check if a merge of $t_a$, $t_b$ is valid  by checking whether $t_a$ can replace $t_b$ in the example strings, and vice-versa. The strings derivable from an arbitrary nonterminal $N$ in \trees{} are the concatenated leaves of the subtree rooted at $N$. We check whether $t_a$ replaces $t_b$ by checking whether the strings produced by replacing strings derivable from $t_a$ by strings derivable from $t_b$, are accepted by the oracle. That is, we take the strings derivable from the trees \trees{}, with holes in place of strings derived from $t_b$. Then we fill the holes with strings derivable by $t_a$. If all the strings are accepted by $\oracle{}$, \ourmethod{} judges the replacement as valid. Section~\ref{sec:string-sampling} details this check and its soundness.

Now the main \ourmethod{} loop starts. From the current  $S$-derived trees \trees{}, \ourmethod{} gets all potential ``bubbles'' for the trees (Algorithm~\ref{alg:full-grammar-learning}, Line~\ref{line:get-bubbles}). For each tree $t\in \trees{}$, \getbubbles{} collects all proper contiguous subsequences of children in $t$. That is, if the tree contains a node $t_i$ with children $C = c_1, c_2, \dots, c_n$, the potential bubbles include all subsequences of $C$ of length greater than one and less than $n$. \getbubbles{} returns all these subsequences as 1-bubbles, and all non-conflicting pairs of these subsequences as 2-bubbles. Two subsequences are non-conflicting if they do not \textit{strictly} overlap: they can be disjoint or one can be a proper subsequence of the other. So $((c_1, c_2, c_3), (c_2, c_3, c_4))$ conflict, but $((c_1, c_2, c_3), (c_2, c_3))$ and $((c_1, c_2, c_3), (c_4, c_5))$ do not. The order in which \ourmethod{} explores these bubbles is important for efficiency; we discuss this further in Section~\ref{sec:bubble-order}.

\begin{figure*}
	\centering
	\small
	\begin{minipage}{0.33\textwidth}
		\scalebox{0.83}{\begin{forest} [$t_0$, l sep={0.5cm}, for tree= {s sep={0.5mm}} [\texttt{w}] [\texttt{h}] [\texttt{i}, fill=lightyellow] [\texttt{l}, fill=lightyellow] [\texttt{e}, fill=lightyellow] [\textvisiblespace{}, fill=lightyellow] [\texttt{n}] [\texttt{=}] [\texttt{=}] [\texttt{n}] [\textvisiblespace{}, fill=lightyellow] [\texttt{d}, fill=lightyellow][\texttt{o}, fill=lightyellow] [\textvisiblespace{}, fill=lightyellow] [$t_0$, for tree= {s sep={0.5mm}} [\texttt{s}] [\texttt{k}] [\texttt{i}] [\texttt{p}]]]  
				\node at (current bounding box.south)
				[draw,rectangle, below=.5ex]
				{\large \emph{Tree 1}};
			\end{forest}
		}
	\end{minipage}
	\begin{minipage}{0.3\textwidth}
		\centering
		\hspace{-2em}
		\scalebox{0.83}{%
			\begin{forest} [$t_0$, l sep={0.5cm}, for tree= {s sep={0.5mm}} [\texttt{w}] [\texttt{h}] [\texttt{i}, fill=mint] [\texttt{l}, fill=mint] [\texttt{e}, fill=mint] [\textvisiblespace{}, fill=mint] [$t_1$, for tree= {s sep={0.5mm}} [\texttt{t}] [\texttt{r}] [\texttt{u}] [\texttt{e}]] [\textvisiblespace{}, fill=mint] [\texttt{d}, fill=mint][\texttt{o}, fill=mint] [\textvisiblespace{}, fill=mint] [$t_0$, for tree= {s sep={0.5mm}} [\texttt{s}] [\texttt{k}] [\texttt{i}] [\texttt{p}]]]  
				\node at (current bounding box.south)
				[draw,rectangle, below=.5ex]
				{\large\emph{Tree 2}};
			\end{forest}
		}
	\end{minipage}
	\begin{minipage}{0.35\textwidth}
		\hspace{-2em}
		\scalebox{0.83}{%
			\begin{forest}
				[$t_0$, l sep={0.5cm}, for tree= {s sep={0.5mm}} [\texttt{i}, fill=mint] [\texttt{f}, fill=mint]  [\textvisiblespace{}, fill=mint] [$t_1$, [\texttt{f}] [\texttt{a}] [\texttt{l}] [\texttt{s}] [\texttt{e}]] [\textvisiblespace{}, fill=mint] [\texttt{t}, fill=mint] [\texttt{h}, fill=mint] [\texttt{e}, fill=mint] [\texttt{n}] [\textvisiblespace{}] [$t_0$, [\texttt{s}] [\texttt{k}] [\texttt{i}] [\texttt{p}]]
				[\texttt{e}] [\texttt{l}] [\texttt{s}] [\texttt{e}] [\textvisiblespace{}] [$t_0$, [\texttt{s}] [\texttt{k}] [\texttt{i}] [\texttt{p}]]]  
				\node at (current bounding box.south)
				[draw,rectangle, below=.5ex]
				{\large\emph{Tree 3}};
			\end{forest}
		}
	\end{minipage}
	\vspace{-0.5em}
	\caption{Partial parse tree $\trees{}$ during run of \ourmethod{} on while, with guide examples ``\texttt{while n==n do skip}'', ``\texttt{if false then skip else skip}'' and ``\texttt{while true do skip}''. 
		\ourmethod{} has applied the 1-bubble ``\texttt{skip}'', which merged with $t_0$, and the 2-bubble (``\texttt{false}'', ``\texttt{true}''). The 4-contexts for ``\texttt{n == n}''' are \hlc[lightyellow]{highlighted in yellow}, and for  $t_1$ are \hlc[mint]{highlighted in green}.
		\vspace{-1em}    
	}
	\label{fig:context}
\end{figure*}

Then, for each potential bubble, \ourmethod{} tries applying it to the existing set of trees \trees{}. Suppose we have a 1-bubble consisting of the subsequence $c_i, c_{i+1}, \dots, c_j$. To apply this bubble,  we replace any sequence of siblings $t_{c_i}, t_{c_{i+1}}, \dots, t_{c_j}$ with labels $c_i, c_{i+1}, \dots, c_j$ in the tree with a new subtree $t_{new} \to t_{c_i}, t_{c_{i+1}}, \dots, t_{c_j}$. Fig.~\ref{fig:motivate-single-bubbles} shows two such bubblings: \texttt{hile} is bubbled into the nonterminal $t_1$ at the top, and \texttt{(n+n)} is bubbled to $t_2$ on the bottom. If the bubbled nodes have structure under them, that structure is maintained: e.g., the bubbling of $t_7\textvisiblespace\texttt{\&}\textvisiblespace t_7$ into $t_9$ at (4) in Fig.~\ref{fig:motivate-full-run}. For a 2-bubble, the same process is repeated for the two subsequences involved.

After applying the bubble, \ourmethod{} checks whether it should be accepted (Line~\ref{line:check-bubbles}). 
Section~\ref{sec:bubble-accepting} formalizes \acceptbubble{}, but essentially, \acceptbubble{} accepts a bubble if the new nonterminals introduced in its application can be validly merged with some other nonterminal node in the tree. 

If the new bubbled nonterminal allows a valid merge with some other nonterminal, \acceptbubble{} returns \texttt{True} as well as the trees with the merge applied (Line~\ref{line:check-bubbles}). We update the best trees \trees{}  to reflect the successful merge (Line~\ref{line:give-merged-tree}), and \getbubbles{} is called again on the new \trees{}. If the bubble is not accepted, \ourmethod{} continues to check the next bubble returned by \getbubbles{} (Line~\ref{line:bubble-loop}).  

The algorithm terminates when none of the bubbles are accepted, i.e. when the trees $\trees{}$ cannot be further generalized, and returns the grammar $\mathcal{G}$ induced by the trees $\trees{}$ (Line~\ref{line:ret-main}).

We can guarantee the following about \ourmethod{} as long as merges are sound,  once we consider the notion of \emph{partially merging} two nonterminals, discussed in  Section~\ref{sec:partial-merges}. 

\textbf{\textsc{Existence Theorem:}} There exists a sequence of $k$-bubbles, that, when considered by \ourmethod{} in order, enable \ourmethod{} to return a grammar $\mathcal{G}$ s.t. $\mathcal{L(G) = L(\grammar_\oracle)}$, so long as the input examples $S$ are exercise all rules of $\mathcal{G}$. 

\textbf{Proof Outline:}  The optimal  bubble order always chooses the right-hand-side of some $N\to \alpha_1\cdots\alpha_n$ in $\mathcal{G}$ as the sequence to bubble,   either as 1-bubble if there exists an expansion for $N$ in the trees already, or as a 2-bubble otherwise.  

Refer to Appendix~\ref{sec:proofs} for  formal treatment of this  and the \textbf{\textsc{Generalization Theorem},} which shows that $k$-bubbles monotonically increase the language of the learned grammar.

\subsection{Ordering Bubbles for Exploration}
\label{sec:bubble-order}

As described in paragraph 5) of Section~\ref{sec:motivation} and alluded to above, the order of bubbles impacts the eventual grammar returned by \ourmethod{}. 
%
%
Unfortunately,
the number of orderings of bubbles is exponential. To have an efficient algorithm in practice, we must make sure the algorithm finds the correct order of bubbles early in its exploration of bubble orders. As such, \getbubbles{}  returns bubbles in an order more likely to enable sound generalization of the grammar being learned. 

As described in the prior section, bubble sequences consist of proper contiguous subsequences of children in the current trees \trees{}. We  increase the maximum length of subsequences considered once all bubbles of shorter length do not enable any valid merges. These subsequences (and their pairs) form the base of 1-bubbles (and 2-bubbles) returned by \getbubbles{}. 

Recall that a bubble should be accepted if the bubbled nonterminal(s) can be merged with an existing nonterminal (or each other). Thus, \getbubbles{} should first return those  bubbles that are likely to be mergeable. We leverage the following observation to return bubbles likely to enable merges. Expansions of a given nonterminal often occur in a similar \emph{context}. The $k$-context of a sequence of sibling terminals/nonterminals $s$ in a tree is the tuple of $k$ siblings to the left of $s$ and $k$ siblings to right of $s$. 

Fig.~\ref{fig:context}  shows an example of a run of \ourmethod{} on the while language, after the application of the 1-bubble ``\texttt{skip}'' and the 2-bubble (``\texttt{false}'', ``\texttt{true}''). The set of 4-contexts  for the sequence ``\texttt{n\textvisiblespace{}==\textvisiblespace{}n}'' is $\{((\text{\hlc[lightyellow]{\texttt{i}, \texttt{l}, \texttt{e}, \textvisiblespace{}}}), (\text{\hlc[lightyellow]{\textvisiblespace{}, \texttt{d}, \texttt{o}, \textvisiblespace{}}}))\}$. Similarly, ``$t_1$'''s 4-contexts are  $\{((\text{\hlc[mint]{\textit{S}, \texttt{i}, \texttt{f}, \textvisiblespace{}}}), (\text{\hlc[mint]{\textvisiblespace{}, \texttt{t}, \texttt{h}, \texttt{e}}})), 
((\text{\hlc[mint]{\texttt{i}, \texttt{l}, \texttt{e}, \textvisiblespace{}}}), (\text{\hlc[mint]{\textvisiblespace{}, \texttt{d}, \texttt{o}, \textvisiblespace{}}}))\}$; ``$S$'' is a dummy  element indicating the start of the example string. Note that ``\texttt{n==n}'' and ``$t_1$'' share the 4-context $\{((\text{{\texttt{i}, \texttt{l}, \texttt{e}, \textvisiblespace{}}}), (\text{{\textvisiblespace{}, \texttt{d}, \texttt{o},\textvisiblespace{}}}))\}$

With this in mind, \getbubbles{} orders the bubbles in terms of their \emph{context similarity}. Given two contexts $c_0=\left(l_0, r_0\right)$ and $c_1=\left(l_1, r_1\right)$, where $l_i =(l_i^k, l_i^{k-1}, \dots, l_i^0)$
and $r_i =(r_i^0, \dots,r_i^{k-1}, r_i^k )$, we have
$\textit{contextSim}(c_0, c_1) = \textit{kTupleSim}(l_0, l_1) + \textit{kTupleSim}(r_0, r_1)$,
where
$$
\textit{kTupleSim}(t_0, t_1) = \begin{cases}
\frac{1}{2}\text{ if } t_0 = t_1 \\
\sum_{i=0}^k\frac{\mathbb{1}_{=}(t_0^i, t_1^i)}{2^{i+2}}
\end{cases}
$$
where $\mathbb{1}_{=}$ is the indicator function, returning  1 if its arguments are equal and 0 otherwise. This similarity function gives most weight to the context elements closest to the bubble. 

With this in mind, we define set context similarity as the maximum similarity of two contexts within the set:
\vspace{-0.25em}
$$
\textit{setContextSim}(C_0, C_1)= \max_{c_0\in C_0, c_1\in C_1}{\textit{contextSim}(c_0,c_1) }.
$$
\vspace{-0.25em}
In our running example, the context similarity is 1 because  \texttt{n==n}'s 4-context set is a subset of $t_1$'s 4-context set. 

To form bubbles, \getbubbles{} first traverses all the trees \trees{} currently maintained by \ourmethod{}. It considers each proper contiguous subsequence of siblings in the trees. For each subsequence $s$, it collects the $k$-contexts for $s$, as well as the occurrence count of the subsequence $\textit{occ}(s)$. In Fig.~\ref{fig:context}, $\textit{occ}(\texttt{while})=2$, $\textit{occ}(t_1)=2$ and $\textit{occ}(\texttt{n\textvisiblespace{}==\textvisiblespace{}n})=1$. In our implementation we take $k=4$.

\ourmethod{} then creates a 2-bubble for each pair of sequences $(s_1, s_2)$  where both $|s_1|>1$ and $|s_2|>1$. The similarity score of this 2-bubble is $\textit{setContextSim}(\textit{contexts}(s_1),$ $\textit{contexts}(s_2))$ and its frequency score is the average frequency of the two sequences in the bubble $\frac{\textit{occ}(s_1) + \textit{occ}(s_2)}{2}$.
Additionally, for each sequence $s_0$ with $|s_0| > 1$, \ourmethod{} creates a 1-bubble $(s_0)$. Let $S_1$ be the set of length-one subsequences. The similarity score of $(s_0)$ is $\max_{s_1\in S_1}\textit{setContextSim}(\textit{contexts}(s_0),$ $ \textit{contexts}(s_1))$ and its frequency score is $\textit{occ}(s_0)$. 

Finally, \getbubbles{} takes the top-$n$ bubbles as sorted primarily by similarity, and secondarily by frequency.  Intuitively, high-frequency sequences may correspond to tokens in the oracle's language.  The order of bubbles is shuffled to prevent all runs of \ourmethod{} from getting struck in the same manner. We find $n=100$ to be effective in practice.

\subsection{Accepting Bubbles}
\label{sec:bubble-accepting}

\begin{figure}
    \begin{center} 
    \small
     \begin{forest}
[$t_0$, l sep={0.2cm}, for tree= {l sep = {0.2cm}, s sep={0.5mm}} 
    [\texttt{i}] 
    [\texttt{f}]  
    [\textvisiblespace{}] 
    [$t_1$, [\texttt{n}, fill=lightblue] [\textvisiblespace{}] [\texttt{=}] [\texttt{=}] [\textvisiblespace{}] [\texttt{n}, fill=lightyellow]]
    [\textvisiblespace{}] 
    [\texttt{t}] 
    [\texttt{h}] 
    [\texttt{e}] 
    [\texttt{n}, fill=mint] 
    [\textvisiblespace{}] 
    [$t_0$, [\texttt{L}][\textvisiblespace{}] [\texttt{=}] [\textvisiblespace{}][\texttt{n}, fill=orange]]
    [\texttt{e}] 
    [\texttt{l}] 
    [\texttt{s}] 
    [\texttt{e}] 
    [\textvisiblespace{}] 
    [$t_0$, [\texttt{L}] [\textvisiblespace{}][\texttt{=}] [\textvisiblespace{}] [\texttt{n}, fill=orange]]
]  
    \end{forest}
    \end{center}
    \vspace{-1em}
    \boxalign{
    \vspace{0.2cm}
    \small
        \begin{align}
        \begin{split}
    &t_0 \to~ \texttt{i f \textvisiblespace{}}~t_1~\texttt{\textvisiblespace{} t h e }\mathcolorbox{mint}{t_\texttt{n}}\textvisiblespace{}~t_0~\texttt{\textvisiblespace{} e l s e}\\
    &t_1  \to~ \mathcolorbox{lightblue}{t_\texttt{n}}\texttt{\textvisiblespace{}==\textvisiblespace{}}\mathcolorbox{lightyellow}{t_\texttt{n}}\\
    &t_0  \to~ {\texttt{L}}\texttt{\textvisiblespace{}=\textvisiblespace{}}\mathcolorbox{orange}{t_\texttt{n}}\\
    & t_\texttt{n} \to \texttt{n}
    \end{split}
    \end{align}
    }    
    \vspace{-1em}
   
    \boxalign{
    \vspace{0.2cm}
    \small
        \begin{align}
        \begin{split}
    &t_0 \to~ \texttt{i f \textvisiblespace{}}~t_1~\texttt{\textvisiblespace{} t h e }\mathcolorbox{mint}{t_{n_1}}\textvisiblespace{}~t_0~\texttt{\textvisiblespace{} e l s e}\\
    &t_1  \to~ \mathcolorbox{lightblue}{t_{n_2}}\texttt{\textvisiblespace{}==\textvisiblespace{}}\mathcolorbox{lightyellow}{t_{n_3}}\\
    &t_0  \to~ {\texttt{L}}\texttt{\textvisiblespace{}=\textvisiblespace{}}\mathcolorbox{orange}{t_{n_4}}\\
    & t_{n_1} \to \texttt{n}\phantom{aaa}t_{n_2} \to \texttt{n}\phantom{aaa}t_{n_3} \to \texttt{n}\phantom{aaa}t_{n_4}\to \texttt{n}
    \end{split}
    \end{align}%
\vspace{-1em}}
   \vspace{-0.5em}
    \caption{Example tree and rules in its induced grammar which have $t_\texttt{n}$ in their expansion (1), and the same grammar with $t_\texttt{n}$ split at different positions.  
 For simplicity,  nonterminals of the form $t_c\to c$---other than $t_\texttt{n}$ in (1)---are collapsed to $c$.}
 
 \vspace{-1em}
    \label{fig:partial}
\end{figure}
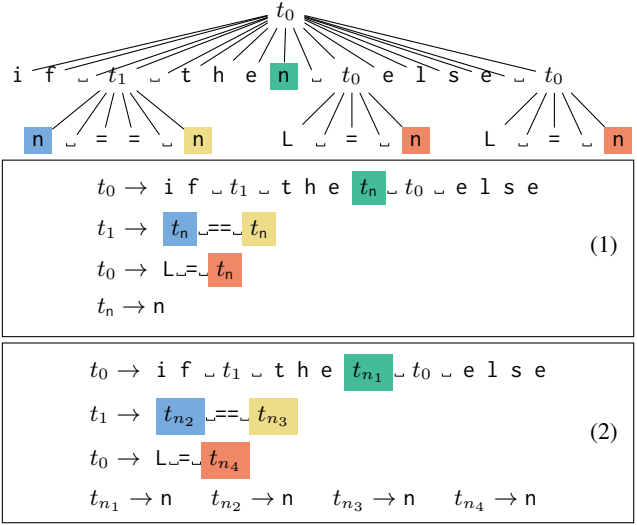

The second key component of \ourmethod{} is deciding whether a given bubble should be accepted: this section formalizes how \acceptbubble{} works. 
At the core of \acceptbubble{} is the concept of whether two labels $t_a,t_b$ can be merged. We say that $t_a$ and $t_b$ can be merged, i.e. $\merges{}(t_a, t_b)$, if and only if $\replaces{}(t_a, t_b)$---that is, all occurrences of $t_b$ can be replaced by $t_a$ in the grammar---and  $\replaces{}(t_b, t_a)$. We formalize how $\replaces{}$ is checked in the next section.  

\subsubsection{2-Bubbles} \ourmethod{} accepts a 2-bubble $(s_1, s_2)$ with labels $t_{s_1},t_{s_2}$ only if  $\merges{}(t_{s_1},t_{s_2})$. 
Intuitively, this is because both bubbles should be kept only if they together expand the grammar. For example, suppose we apply the 2-bubble (``\texttt{n\textvisiblespace{}==\textvisiblespace{}n}'', ``\texttt{lse}'') to the trees in Fig.~\ref{fig:context}, resulting in nonterminals $t_{n==n}\to\texttt{\texttt{n\textvisiblespace{}==\textvisiblespace{}n}}$  and $t_{lse}\to\texttt{lse}$. While $t_{n==n}$ can merge with $t_1$,  $t_{lse}$ does not contribute to this merging.  So, (``\texttt{n\textvisiblespace{}==\textvisiblespace{}n}'') should be accepted only as a 1-bubble.

\subsubsection{1-Bubbles} 
\label{sec:partial-merges}
Recall that \ourmethod{} scores 1-bubbles highly if they are likely to merge with an existing nonterminal. Let $\textit{NTs}(\trees)$ be the nonterminal labels present in the current set of trees $\trees{}$. Given a 1-bubble $(s_1)$ with label $t_{s_1}$,  we go through each $t_i\in \textit{NTs}(\trees)$ and check whether $\merges{}(t_i, t_{s_1})$. If $\merges{}(t_i, t_{s_1})$ is true for some $t_i\in \textit{NTs}(\trees)$, then $\acceptbubble{}$ accepts the bubble $(s_1)$. 

However, if $t_{s_1}$ cannot merge with any $t_i\in \textit{NTs}(\trees)$, \ourmethod{} also looks for \emph{partial merges}. Partial merging works as follows. Let $\textit{CNTs}(\trees)$ be the \emph{character nonterminal} labels present in the current set of trees $\trees{}$. A \emph{character nonterminal} is a nonterminal whose expansions only of a single terminal element, e.g., $t_\texttt{n}\to \texttt{n}$ or $t_1\to \texttt{1}\;|\;\texttt{2}\;|\;\texttt{3}$. 

For each $t_c\in\textit{CNTs}(\trees)$, the partial merging algorithm identifies all the different occurrences of $t_c$ in the right-hand-side of  expansions in \trees{}'s induced grammar. For instance, in the grammar fragment (1) of Fig.~\ref{fig:partial}, we see the nonterminal $t_\texttt{n}$, corresponding to ``\texttt{n}'', occurs 4 distinct times in right-hand-sides of expansions. The partial merging algorithm then modifies the grammar so that the $i^{th}$ occurrence of $t_c$ is replaced with a fresh nonterminal $t_{c_i}$. Each $t_{c_i}$ expands to the same bodies as $t_c$; i.e. $t_{c_i} \to c$. This replacement process is illustrated in the grammar fragment  (2) of Fig.~\ref{fig:partial}: the four occurrences of $t_\texttt{n}$ have been replaced with $t_{n_1}$, $t_{n_2}$, $t_{n_3}$, and $t_{n_4}$. Finally,  we get to the \textit{merging} in \textit{partial merging}: for each $t_{c_i}$, the algorithm checks if $\merges{}(t_{c_i}, t_{s_1})$. If $\merges{}(t_{c_i}, t_{s_1})$ for any $t_{c_i}$, \ourmethod{} accepts the bubble $(s_1)$, and $t_{s_1}$ is merged with all such $t_{c_i}$. The $t_{c_j}$ which cannot be merged with $t_{s_1}$ are restored to the original  nonterminal $t_c$. 

The term partial merge refers to the fact that we have effectively merged $t_{s_1}$ with \emph{some} of the occurrences of $t_c$ in rule expansions. This step is useful when \ourmethod{}'s initial trees---which map each character to a single nonterminal---use the same nonterminal for characters that are conceptually separate.  For instance, consider the 1-bubble (\texttt{(n+n)}), with label $t_{(n+n)}$. Given the tree in Fig.~\ref{fig:partial}, $\merges{}(t_\texttt{n}, t_{(n+n)})$ fails because ``\texttt{(n+n)}'' cannot replace the ``\texttt{n}'' in ``\texttt{then}''. In fact, $t_{(n+n)}$ cannot merge with any $t_i\in \textit{NTs}(\trees)$ initially. But the partial merge process splits $t_\texttt{n}$ into $t_{n_1}$, $t_{n_2}$, $t_{n_3}$, $t_{n_4}$, and \ourmethod{} finds that $t_{(n+n)}$ in fact merges with $t_{n_2}$, $t_{n_3}$ and $t_{n_4}$. So, it is merged with those nonterminals and accepted. 

\textit{Note}: though we consider only partial merges on character nonterminals for efficiency reasons,  the concept of partial merging can be applied to any pair of nonterminals.

In summary, a 1-bubble $(s_1)$ with label $t_{s_1}$ is accepted if either: (1) for some $t_i\in\textit{NTs}(\trees)$, $\merges{}(t_i, t_{s_1})$, or (2) for some $t_c\in\textit{CNTs}(\trees)$, $t_{s_1}$ can be partially merged with $t_c$.

\subsection{Sampling Strings for Replacement Checks}
\label{sec:string-sampling}

The final important element affecting the performance of \ourmethod{} is how exactly we determine whether the merge of two nonterminals labels is valid. 
Recall that $\merges{}(t_a, t_b)$ if and only if $\replaces{}(t_a, t_b)$ and $\replaces{}(t_b, t_a)$.

We implement $\replaces{}(t_\text{replacer}, t_\text{replacee})$ as follows. From the current parse trees, we derive the \textit{replacee strings}: the strings derivable from the parse trees in \textit{trees}, but with holes instead of the strings derived from $t_\text{replacee}$. Then, we derive a set of \textit{replacer strings}: the strings derivable from $t_\text{replacer}$ in the trees. Finally, we create the set of \emph{candidate strings}  by replacing the holes in the replacee strings with the replacer strings. If \oracle{} rejects any candidate string, the merge is rejected, and  \replaces{} returns false.

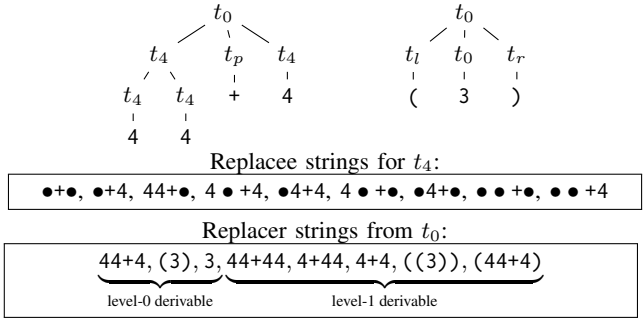
\begin{figure}
	\centering
	\small
	\begin{forest} baseline
		[$t_0$,  for tree={l sep=1mm, l=1mm}
		[$t_4$, [$t_4$, [\texttt{4}]]  [$t_4$, [\texttt{4}]]]
		[$t_p$, [\texttt{+}]]
		[$t_4$, [\texttt{4}]]
		] 
	\end{forest}
	\hspace{1cm}
	\begin{forest} baseline
		[$t_0$, for tree={l sep=1mm, l=1 mm}
		[$t_l$, [\texttt{(}]] 
		[$t_0$, [\texttt{3}]]
		[$t_r$, [\texttt{)}]]
		] 
	\end{forest}
	
	Replacee strings for $t_4$:\\
	\noindent\fbox{
		\parbox{0.9\columnwidth}{\small \centering $\bullet\texttt{+}\bullet$, $\bullet\texttt{+4}$, $\texttt{44+}\bullet$, $\texttt{4}\bullet\texttt{+4}$, $\bullet\texttt{4+4}$, $\texttt{4}\bullet\texttt{+}\bullet$, $\bullet\texttt{4+}\bullet$, $\bullet\bullet\texttt{+}\bullet$, $\bullet\bullet\texttt{+4}$}
	} 
	
	
	\vspace{0.5em}
	Replacer strings from $t_0$:\\
	\noindent\fbox{
		\parbox{0.9\columnwidth}{\centering $\small\underbrace{\texttt{44+4},\texttt{(3)},  \texttt{3},}_\text{level-0 derivable} \underbrace{\texttt{44+44}, \texttt{4+44}, \texttt{4+4},  \texttt{((3))}, \texttt{(44+4)}}_\text{level-1 derivable}$}
	}
	\vspace{0.5em}
	
	\caption{Two partial parse trees and examples of replacee and replacer strings. The symbol $\bullet$ designates holes which will be replaced by (level-$n$ derivable) replacer strings.}
	\label{fig:replacer-replacee-strings}
	
	\vspace{-1.5em}
\end{figure}

Fig.~\ref{fig:replacer-replacee-strings} shows how replacer and replacee strings are computed in the call to $\replaces{}(t_\text{0}, t_\text{4})$, i.e. whether $t_0$ can replace $t_4$. Replacee strings for a node in the parse tree are computed by taking the product of replacee strings for all its children; the nonterminal being replaced becomes a hole. 

Level-0 replacer strings for $t_i$ are just the strings that directly derivable from $t_i$ in the tree; in Fig.~\ref{fig:replacer-replacee-strings}, the level-0 derivable strings of $t_0$ are \texttt{44+4},  \texttt{(3)},  \texttt{3}, and the level-0 derivable strings of $t_4$ are \texttt{44},  \texttt{4}. Then, the set of level-$n$ derivable strings for a node is the set derived from taking the product of all level-($n-1$) derivable strings for each child of a node. The level-1  replacer strings for $t_0$ are shown in Fig.~\ref{fig:replacer-replacee-strings}.


When \replaces{} is run in the full \textsc{MergeAllValid} call or while evaluating a 1-bubble, we use only level-0 replacer strings. However, we found that level-1 replacer strings greatly increased soundness at a low runtime cost for 2-bubbles. Intuitively this is because nonterminals from new bubbles tend to have less structure underneath them than existing nonterminals in the trees. So it is faster to compute level-1 replacer strings for these new bubble-induced nonterminals. 

Note that the both the number of replacee strings and of level-n derivable replacer strings grows exponentially. So, instead of taking the entire set of strings derivable in this manner, if there are more than $p$ of them, we uniformly sample $p$ of them. In our implementation we use $p=50$, to make the number of parse calls reasonable in terms of runtime. 

Unfortunately, this process allows  unsound merges, where all candidate strings are accepted by the oracle, but the merge adds oracle-invalid inputs to the language of the learned grammar. First, because only $p$ candidates are sampled. Second, because the replacee strings are effectively ``level 0'', and thus, not reflective of the current induced grammar from the trees. 
Third, 
because a candidate string is produced by replacing all its holes with a single replacer string, rather than filling holes with different replacer strings.  Taking $p\to\infty$,  
$n\to\infty$ for the level-$n$ replacer strings,  and filling different holes with different replacer strings would ensure  sound merges.

\subsection{Pre-tokenization}
\label{sec:pre-tokenization}
Since \ourmethod{} considers 2-bubbles, it is effectively $n^4$ in the total length of examples $n$.  So, to improve performance as $n$ gets large and reduce the likelihood of creating ``breaking'' bubbles, in our implementation we use a simple heuristic to pre-tokenize the values at leaves, rather than considering each character as a leaf. We group together sequences of contiguous characters of the same class (lower-case, upper-case, whitespace, digits) into leaf tokens. Punctuation and non-ASCII characters are still treated as individual characters. We then run the \ourmethod{} as described previously. To ensure generalization, we add a last stage which tries to expand these tokens into the entire character class: e.g. if $t_1 \to \texttt{abc} | \texttt{cde}$, we check whether $t_1$ can be replaced by any sequence of lower-case letters, letters, or alphanumeric characters. We construct the replacee strings as described above, and sample 10 strings from the expanded character classes as replacer strings.


\section{Evaluation}
\label{sec:evaluation}

\begin{figure*}
	\begin{subfigure}{0.3\columnwidth}
		\centering
		\includegraphics[width=\columnwidth]{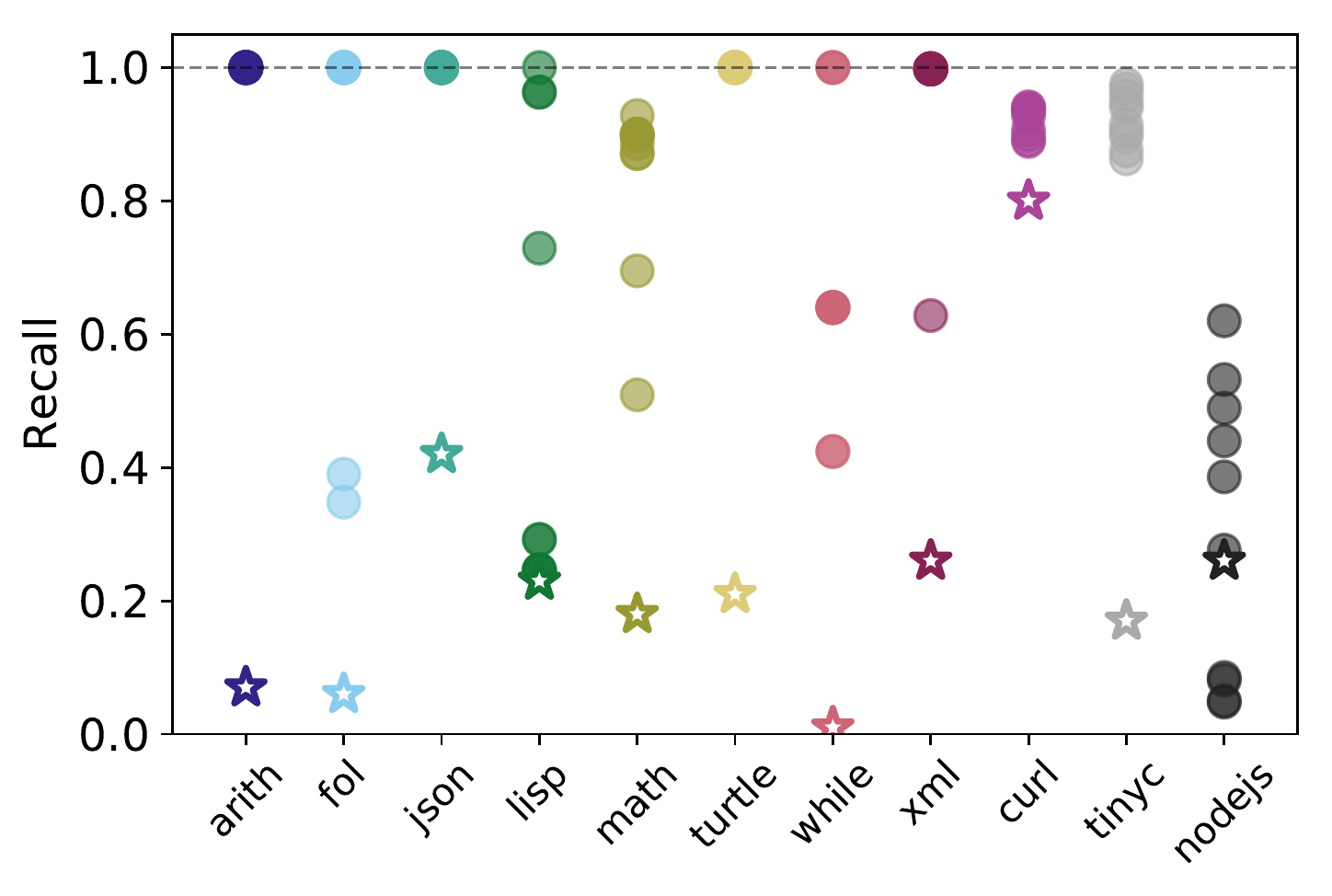}
		\vspace{-2em}
		\caption{Recall. Higher is better.}
		\label{fig:recall-scatter}
	\end{subfigure}%
	\begin{subfigure}{0.3\columnwidth}
		\centering
		\includegraphics[width=\columnwidth]{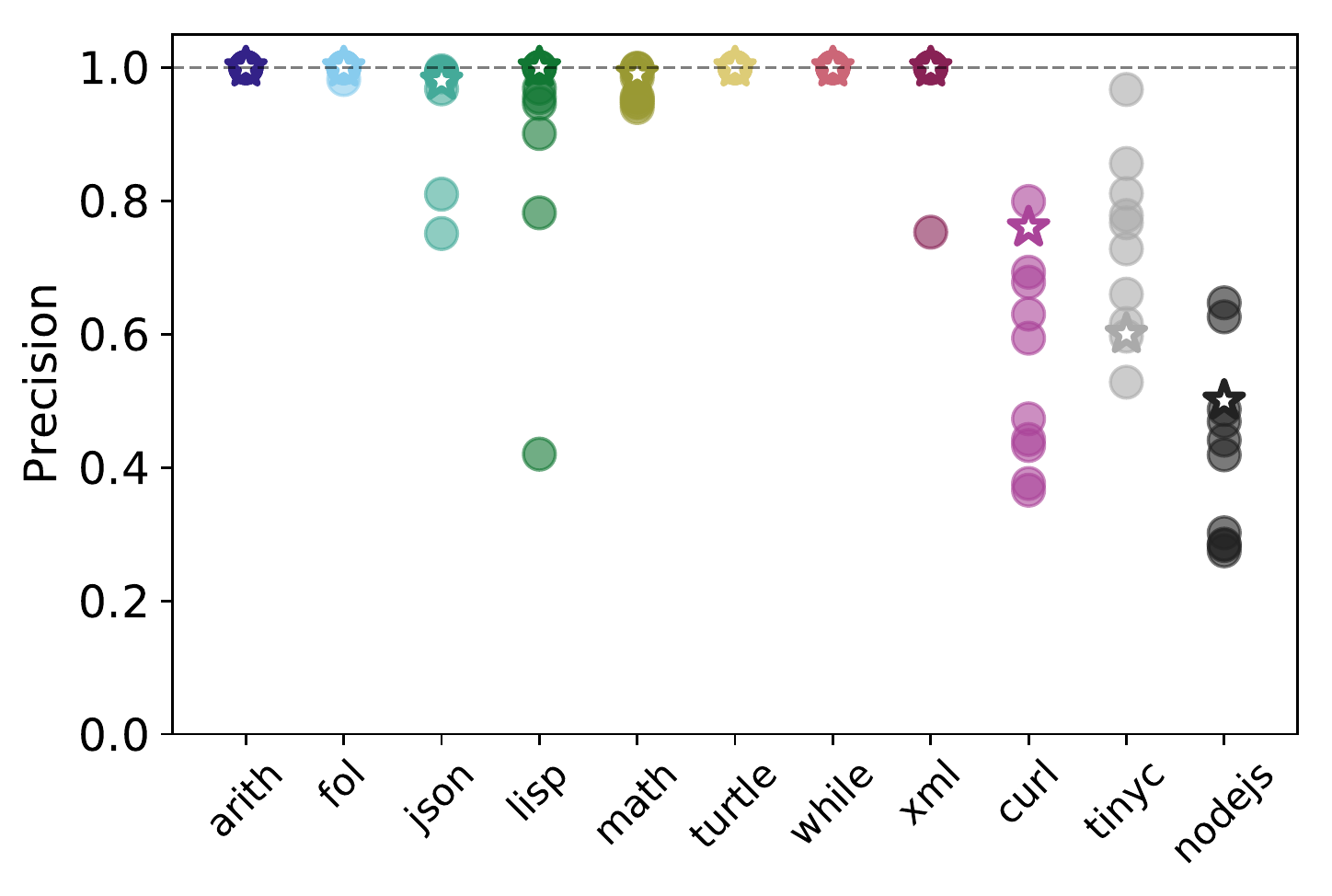}
		\vspace{-2em}
		\caption{Precision. Higher is better.}
		\label{fig:precision-scatter}
	\end{subfigure}
	\begin{subfigure}{0.3\columnwidth}
		\centering
		\includegraphics[width=\columnwidth]{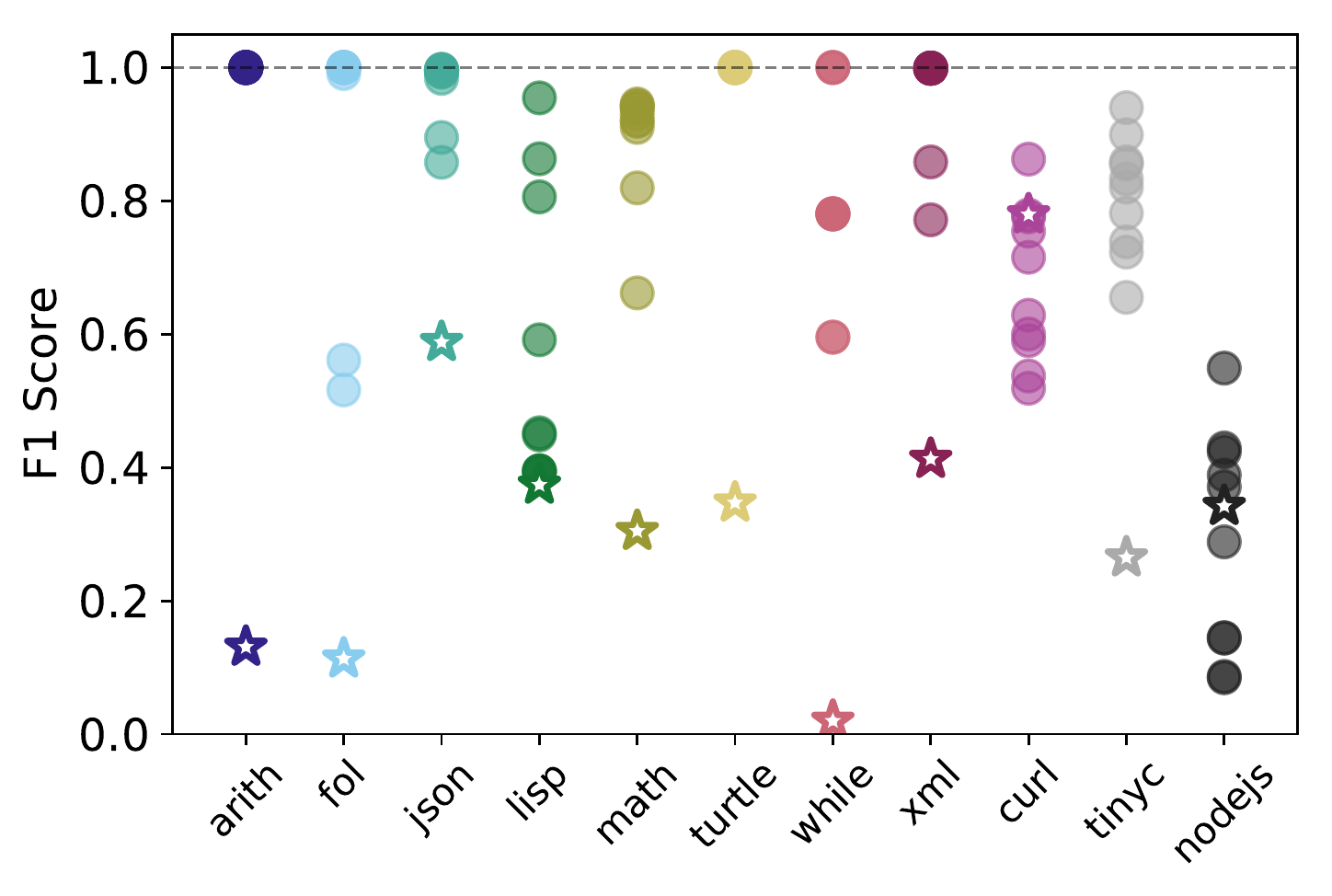}
		\vspace{-2em}
		\caption{F1 Score. Higher is better.}
		\label{fig:f1-scatter}
	\end{subfigure}
	\vspace{-0.5em}
	\caption{Recall, precision, and F1 score for each of the 10 runs of \ourmethod{} (plotted with $\bullet$) and GLADE (plotted with \ding{73}). }
	\label{fig:precision-recall-scatter}
\end{figure*}

\renewcommand{\arraystretch}{0.9}
\setlength{\tabcolsep}{3pt}
\begin{figure*}[t] \TopFloatBoxes
	\begin{floatrow}
		\centering
		\ttabbox[0.695\textwidth]{%
			\vspace{-1em}
			\resizebox{0.695\textwidth}{!}{%
			 \begin{tabular}{lcccccrccccc}
			 \toprule
			           & \multicolumn{5}{c}{\ourmethod{}} & \phantom{a} & \multicolumn{5}{c}{GLADE} \\
			           \cmidrule{2-6} \cmidrule{8-12} \vspace{-0.7em}\\
			 \textbf{Bench.}\phantom{ab} & Recall & Precision & F1 Score & Time(s) & \#\,Queries &&
			 R & P & F1 & Time(s) & \#\,Queries
			 \\
			 \midrule
			 arith & \textbf{1.00} $\pm$ 0.00 & 1.00  $\pm$ 0.00& \textbf{1.00} $\pm$ 0.00& 3 $\pm$ 0& 828 $\pm$ 37&& 0.07 & 1.00 & 0.13 & 12 & 2.3K \\
			  fol & \textbf{0.87} $\pm$ 0.25 & 1.00 $\pm$ 0.01& \textbf{0.91} $\pm$ 0.18& 372 $\pm$ 36& 33K$\pm$ 3.7K&&  0.06 & 1.00 & 0.11 & 107 & 20K\\ 
			 json & \textbf{1.00}  $\pm$ 0.00 & 0.95  $\pm$ 0.08 & 0.97  $\pm$ 0.05 & 76  $\pm$ 11 & 16K  $\pm$ 1K && 0.42 & 0.98 & 0.59 & 61 & 11K \\
			 lisp &\textbf{0.52}  $\pm$ 0.33& 0.90  $\pm$ 0.17 & 0.57  $\pm$ 0.21 & 16  $\pm$ 4 & 3.6K $\pm$ 603 && 0.23 & 1.00 & 0.38 & 20 & 3.8K\\
			 math. & \textbf{0.84}   $\pm$ 0.12 & 0.97 $\pm$ 0.02  & \textbf{0.89}  $\pm$ 0.08 & 65  $\pm$ 6 & 11K $\pm$ 1.1K &&  0.18 & 0.99 & 0.31 & 103 & 19K\\
			 turtle &\textbf{1.00}  $\pm$ 0.00 & 1.00  $\pm$ 0.00 & \textbf{1.00}  $\pm$ 0.00& 84  $\pm$ 8 & 10K  $\pm$ 1.1K &&0.21 & 1.00 & 0.34 & 75 & 14K  \\
			 while & \textbf{0.70}  $\pm$ 0.21 & 1.00  $\pm$ 0.00 & \textbf{0.81}  $\pm$ 0.14 & 54  $\pm$ 5 & 13K $\pm$ 1.5K && 0.01 & 1.00 & 0.02 & 50 & 9.1K \\
			  xml &\textbf{0.96}   $\pm$ 0.11 & 0.98  $\pm$ 0.07 & \textbf{0.96}  $\pm$ 0.08 & 205  $\pm$ 34 & 14K $\pm$ 2.4K && 0.26 & 1.00 & 0.42 & 81& 15K
			  \\
			  \cmidrule{2-12}
			 curl & 0.92  $\pm$ 0.02& 0.55  $\pm$ 0.14 & 0.68  $\pm$ 0.11 & 111  $\pm$ 12 & 25K  $\pm$ 3.1K && 0.80 & 0.76 & 0.78 & 112 & 30K  \\
			 tinyc & \textbf{0.92}  $\pm$ 0.04 & 0.73  $\pm$ 0.13 & \textbf{0.81}  $\pm$ 0.08 & 6.4K $\pm$ 1.2K & 112K  $\pm$ 32K && 0.17 & 0.60 & 0.26 & 917 & 252K\\
			 nodejs & 0.30  $\pm$ 0.21 & 0.42  $\pm$ 0.13 & 0.29  $\pm$ 0.16 & 46K  $\pm$ 22K & 142K  $\pm$ 90K& & 0.26 & 0.50 & 0.34 & 38K & 113K \\
			 \bottomrule
			 \end{tabular}
		}
		}{%
			\caption{Summary of results for \ourmethod{} and GLADE. ``R'' is recall, ``P'' is precision. Results for \ourmethod{} are listed as the means over 10 runs with $\pm$ the standard deviation. Bolded results are 2$\times$ better.} 
			\label{table:perf-results}}
		\ttabbox[\Xhsize]{%
			\vspace{-1em}
			\resizebox{0.28\textwidth}{!}{%
			\begin{tabular}{lccc}
				\toprule
				& \multicolumn{3}{c}{CLGen LSTM}  \\
				\cmidrule{2-4}
				\textbf{Bench.} & Time(s) & Model Time(s) & Precision \\ 
				\midrule
				{arith}     & 172                     & 9          & 0.002     \\
				{fol}       & 177                     & 12         & 0.460      \\
				{json}      & 178                     & 11         & 0.625     \\
				{lisp}      & 173                     & 9          & 0.367     \\ 
				{mathexpr}  & 176                     & 12         & 0.393      \\
				{turtle}    & 176                     & 10        & 0.367     \\
				{while}     & 167                     & 9          & 0.012     \\
				{xml}       & 171                     & 12         & 0.228     \\
							  				\cmidrule{2-4}
				{curl}      & 176                     & 12         & 0.434     \\
				{tinyc}     & 189                     & 21         & 0.062     \\
				{nodejs}    & 176                     & 18        & 0.111     \\
				\bottomrule
			\end{tabular}
		}
} {%
\caption{Results for CLGen's core LSTM \cite{Cummins17}. ``Model Time'' is the logged model training time.}
\label{table:lstm-results}
}
\end{floatrow}
\end{figure*}

We seek to answer the following research questions:
\begin{enumerate}[label=RQ\arabic*., wide=0pt, leftmargin=*]
    \item Do \ourmethod{}'s mined grammars generalize better (have higher recall)  than state-of-the-art?
    \item Do \ourmethod{}'s mined grammars produce more valid inputs (have higher precision)  than state-of-the-art?
    \item How does the nondeterminism in \ourmethod{} cause its behavior to vary across different invocations?
    \item How does \ourmethod{}'s performance compare to that of deep-learning approaches?
    \item What are \ourmethod{}'s major performance bottlenecks? 
    \item What do \ourmethod{}'s mined grammars look like? 
\end{enumerate}

\subsection{Benchmarks} 
We evaluate \ourmethod{} against state-of-the-art blackbox grammar inference tool GLADE~\cite{BastaniGladePLDI2017} on 11 benchmarks.

The first 8 benchmarks consist of an ANTLR4~\cite{ParrANTLRSPE1995} parser for the ground-truth grammar as oracle and a randomly generated set of training examples $S$. $S$ is sampled to cover all of the rules in the ground-truth grammar, while keeping the length of each example $s\in S$ small. The test set is randomly sampled from the ground-truth grammar. Essentially, this ensures that the maximal generalization of $S$ covers the entire test set. Other than \texttt{turtle} and \texttt{while}, these benchmarks come from prior work~\cite{BastaniGladePLDI2017,WuReinamFSE2019,GopoinathMimidFSE2020}:
\begin{itemize}
    \item \textbf{arith}: operations between integers, can be parenthesized
    \item \textbf{fol}: a representation of first order logic, including qualifiers, functions, and predicates
    \item \textbf{json}: JSON with objects, lists, strings with alpha-numeric characters, Booleans, \texttt{null}, integers, and floats
    \item \textbf{lisp}: generic s-expression language with ``\texttt{.}'' cons'ing
    \item \textbf{mathexpr}: binary operations and a set of function calls on integers, floats, constants, and variables
    \item \textbf{turtle}: LOGO-like DSL for Python's turtle
    \item \textbf{while}: simple while language as shown in Fig.~\ref{fig:motivating-example}
    \item \textbf{xml}: supporting arbitrary attributes, text, and a few labels
\end{itemize}

The next 3  benchmarks use as oracle a runnable program, and use a random input generator to create $S$ and the test set. $S$ consists of the first 25 oracle-valid inputs generated by the generator, and the test set of the next 1000 oracle-valid inputs generated. In this case, there is no guarantee that the maximal generalization of $S$ covers the test set. 
\begin{itemize}
    \item \textbf{curl}: the oracle is the curl\cite{curl} url parser. We use the grammar in RFC 1738~\cite{URLRFC} to generate $S$ and test set.
    \item \textbf{tinyc}: the oracle is the parser for tinyc~\cite{tinyc}, a compiler for a subset of C. We use the same golden grammar as in Mimid~\cite{GopoinathMimidFSE2020} to generate $S$ and the test set.
    \item \textbf{nodejs}: the oracle is an invocation of \texttt{nodejs --check}, which just checks syntax~\cite{nodejs}. To generate $S$ and the test set, we use Zest's~\cite{Zest} javascript generator.
\end{itemize}

The average length of training examples in the set $S$ is below 20 for all benchmarks except \texttt{tinyc} (77) and \texttt{nodejs} (58). We adjust the maximum bubble length hyperparameter (ref. Section~\ref{sec:bubble-order}) accordingly: the default is to range from 3 to 10, but on \texttt{tinyc} and \texttt{nodejs} we range from 6 to 20.

\subsection{Accuracy Evaluation}
\label{sec:eval-accuracy}

First, we evaluate the accuracy of \ourmethod{} and GLADE's mined grammars with respect to the ground-truth grammar We ran both \ourmethod{} and GLADE with the same oracle example strings. Three key metrics are relevant here:

\textbf{Recall:} the proportion of inputs from the held-out test set---generated by sampling the golden grammar/generator---that are accepted by the mined grammar. We use a test set size of 1000 for all benchmarks.

\textbf{Precision:} the proportion of inputs sampled from the mined grammar that are accepted by the golden grammar/oracle. We sample 1000 inputs from the mined grammar to evaluate this.

\textbf{F1 Score:} the harmonic mean of precision and recall. It is trivial to achieve high recall but low precision (mined grammar captures any string) or low recall but high precision (mined grammar captures only the string in $S$); F1 measures the tradeoff between the two.

\textit{Results.} As \ourmethod{} is nondeterministic in the order of bubbles explored, we ran it 10 times per benchmark. As GLADE is deterministic, we ran it only once per benchmark. 

Table~\ref{table:perf-results} shows the overall averaged results, Fig~\ref{fig:precision-recall-scatter} the individual runs.
We see from the table that on average, \ourmethod{} achieves higher recall than GLADE \emph{on all benchmarks}, and it achieves higher F1 score on all but 2 benchmarks. \ourmethod{} achieves \textit{over 2$\times$} higher recall on 9 benchmarks, and \textit{over 2$\times$} higher F1 score on 7 benchmarks.

Even for those benchmarks where \ourmethod{} does not have a higher F1 score on average,  Fig.~\ref{fig:f1-scatter} shows that \ourmethod{} outperforms GLADE on some runs. For \texttt{nodejs}, on 5 runs, \ourmethod{} achieves a higher F1 score, ranging from 0.37 to 0.55. 
For \texttt{curl}, on 2 runs \ourmethod{} achieves F1 scores greater than or equal to GLADE's: 0.78 and 0.86. It makes sense that GLADE performs well for \texttt{curl}: the url language is regular, and the first phase of GLADE's algorithm works by building up a regular expressions. 
Nonetheless  Fig.~\ref{fig:recall-scatter} shows that \ourmethod{} achieves consistently higher recall on \texttt{curl}.

Overall, on average across all runs and benchmarks, \ourmethod{} achieves \textbf{4.98$\times$ higher recall} than GLADE, while maintaining $0.96\times$ its precision. So, on our benchmarks, the answer to RQ1 is in the affirmative, while the answer to RQ2 is not. Given that \ourmethod{} still achieves a \textbf{3.13$\times$ higher F1 score} on average, and that higher generalization (in the form of recall) is much more useful if the mined grammar is used for fuzzing, we find this to be a very positive result.

However,  we see from the standard deviations in Table~\ref{table:perf-results} that \ourmethod{}'s performance varies widely on some benchmarks, notable \texttt{fol}, \texttt{lisp}, \texttt{while}, and \texttt{fol}. Fig.~\ref{fig:precision-recall-scatter}, which shows the raw data, confirms this. In Fig.~\ref{fig:recall-scatter}, we see that the performance on the \texttt{lisp} benchmark is quite bimodal. All of the mined grammars with recall around 0.25 fail to learn to cons parenthesized s-expressions. This may be because the minimal example set did not actually have an example of this nesting. On \texttt{nodejs}, the two runs with recall less than 0.1 find barely any recursive structures, suggesting that on larger example sets, \ourmethod{} may get lost in bubble order. Overall, the answer to RQ3 is that \ourmethod{}'s nondeterministic bubble ordering can have very large impacts on the results. We discuss possible mitigations in Section~\ref{sec:discussion}.

\subsection{Comparison to Deep Learning Approaches}
\label{sec:deep-learning-eval}

Recently there has been interest in using machine learning to learn input structures. 
For instance, Learn\&Fuzz trains a seq-2-seq model to model the structure of PDF objects \cite{LearnFuzz}; it uses information about the start and end of pdf objects as well as the importance of different characters in its sampling strategy.  DeepSmith~\cite{Cummins19} trains an LSTM  to model OpenCL kernels for compiler fuzzing, adding additional tokenization and pre-processing stages to CLGen~\cite{Cummins17}.

A natural question is how \ourmethod{} compares to these generative models. We trained the LSTM model from CLGen~\cite{Cummins17}, the generative model behind DeepSmith, on our benchmarks. We removed all the OpenCL-specific preprocessing stages from the pipeline. We used the parameters given as example in the CLGen repo, creating a 2-layer LSTM with hidden dimension 128, trained for 32 epochs. We used  \texttt{\textbackslash n!!\textbackslash n} as an EOF separator. Each sample consisted of 100 characters, split into different inputs where the EOF separator appeared. 

Table~\ref{table:lstm-results} shows the runtime of the model on each benchmark, as well as the precision achieved on the first 1000 samples taken from the model. Generally, we see that the precision is much lower than that of GLADE or \ourmethod{}. On \texttt{arith}, the model over-trains on the EOF separator, adding \texttt{\textbackslash{}n} and \texttt{!} throughout samples. Since the model is generative---it can generate samples but not provide a judgement of sample validity---, we cannot measure Recall as in Table~\ref{table:perf-results}. However, qualitative analysis of the samples suggests  there is not much learned recursive generalization. 
For \texttt{json}, 602 of the 625 valid samples are a single string (e.g., \texttt{"F"}); the other 21 valid samples are numbers,  \texttt{false},  or \texttt{[]}. For \texttt{nodejs},  of the 111 valid samples,  26 are  empty,  24 are a single identifier (e.g. \texttt{a\_0}), 18 are a parenthesized integer or identifier (e.g,. \texttt{(242)}), and 17 are a single-identifier \texttt{throw},  e.g. \texttt{throw (a\_0)}.

These results are not entirely unexpected, because the LSTM underlying CLGen is learning \emph{solely from the input examples}. Both \ourmethod{} and GLADE extensively leverage the oracle, effectively creating new input examples from which to learn. This explains why the runtimes look so different between Tables~\ref{table:perf-results} and~\ref{table:lstm-results}. We see in Table~\ref{table:lstm-results} that the total time to setup and train the model is around 3 minutes for all benchmarks, and the core training time is around 10-20 seconds. We see the model training time is slightly higher for \texttt{tinyc} and \texttt{nodejs}, which had longer input examples. 

Overall, we expect these deep-learning approaches to be more well-suited to a case where an oracle is not available, but large amounts of sample inputs are. These models may also be more reliant input-format specific pre-processing steps, like those used on OpenCL kernels in CLGen and DeepSmith.



\subsection{Performance Analysis}
\label{sec:eval-perf}

\begin{figure}
				\includegraphics[width=0.9\columnwidth]{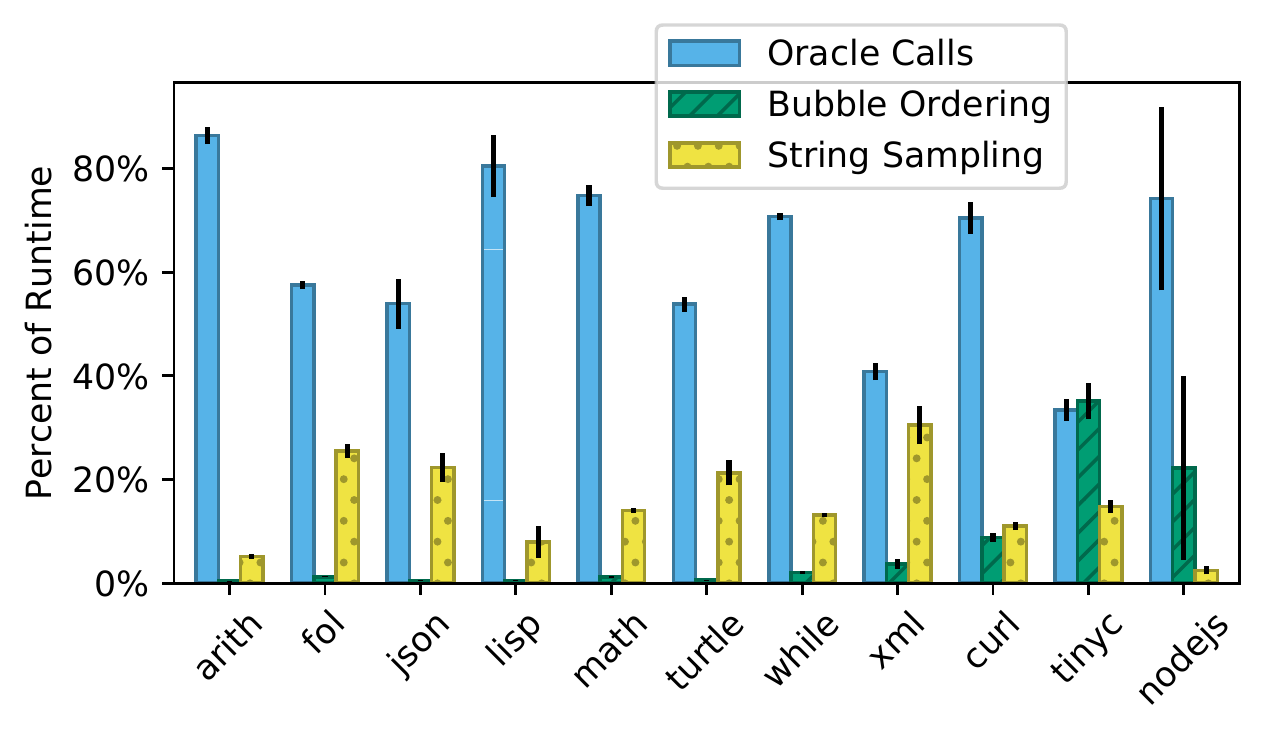}
				\vspace{-1.2em}
				\caption{Average percent of runtime spent in different components of \ourmethod{}. Error bars show std. deviation.}
				\label{fig:perf_distributions}
				
				\vspace{-0.5em}
	\end{figure}

The next question is about \ourmethod{}'s performance. 
Table~\ref{table:perf-results} shows the average \ourmethod{} runtime and number of queries performed for each benchmark, and the same statistics for GLADE. On 7 of 11 benchmarks, \ourmethod{} is on average slower than GLADE; overall across benchmarks, this amounts to an average $1.27\times$ slowdown. This is quite respectable, since \ourmethod{} has a natural runtime disadvantage due to being implemented in Python rather than Java. For the three benchmarks on which \ourmethod{} is over $2\times$ slower than GLADE, it has huge increases in F1 score: $0.11 \to 0.91$ for \texttt{fol}, $0.42 \to 0.96$ for \texttt{xml}, and $0.26\to 0.81$ for \texttt{tinyc}.

The story for oracle queries performed is inversed; \ourmethod{} requires more oracle queries on average on only 4 benchmarks. For all of these except \texttt{nodejs}, \ourmethod{} also had much higher F1 scores. However, \texttt{nodejs} is a benchmark with high variance. On the run with highest F1 score (0.55, higher than GLADE's 0.34), \ourmethod{} takes 86,051 s to run and makes 270k oracle calls. On the fastest run, where \ourmethod{} only gets F1 score 0.14, \ourmethod{} takes 17,775 s and makes 41k oracle calls. That is, the higher performance cost correlates with the slower runs on this benchmark: 5 of the 6 slower runs also have higher F1 scores. 

Overall across all benchmarks, \ourmethod{} performs \textbf{only 0.87$\times$ as many oracle queries as GLADE}. This is encouraging as it gives more room for performance optimizations.
  
Fig.~\ref{fig:perf_distributions} breaks down the average percent of runtime spent in \ourmethod{}'s 3 most costly components: calling the oracle; creating, scoring, and ordering bubbles; and sampling string for replacement checks. The error bars show standard deviation; note the aforementioned high variance for \texttt{nodejs} appears here too. On the minutes-long benchmarks on which \ourmethod{} is at least 10 seconds slower than GLADE, $>20\%$ of the runtime is spent in sampling strings for replacement. The current implementation of this re-traverses the trees $\trees{}$ after each bubble to create these examples.

On the particularly slow benchmarks, \texttt{tinyc} and \texttt{nodejs}, \ourmethod{} spends a long time ordering bubbles. This makes sense because of the larger example length of the benchmarks. It is nonetheless encouraging to see this room for improvement, as \getbubbles{}  re-scores the full set of bubbles each time a bubble is accepted. It should be possible to bring down runtime by only scoring the bubbles that are modified by the application of the just-accepted bubble. On \texttt{nodejs}, \ourmethod{} also spends a long time in oracle queries, because the time for each query is much longer (300 ms vs. 3ms for \texttt{tinyc}). 

Overall, \ourmethod{} has runtime and number of oracle queries comparable with GLADE, while achieving much higher recall and F1 score. As for RQ3, when the length of the examples in $S$ is small, oracle calls dominate runtime. As example length grows, the ordering and scoring of bubbles---particularly computing context similarity---starts to dominate runtime. 

\subsection{Qualitative Analysis of Mined Grammars}

\begin{figure}
\small
\begin{align*}
	\textit{while} \to~& \textit{stmt}\ \term{\textvisiblespace}\ \textit{while} \gramor \term{skip} \gramor \term{L\textvisiblespace{}=\textvisiblespace{}} \\
	\textit{stmt} \to~& \term{while\textvisiblespace}\ \textit{bool}\ \textit{and-space}\ \term{do} \gramor \textit{while}\ \term{\textvisiblespace ;} \\
	&\gramor \term{if\textvisiblespace}\ \textit{bool}\ \textit{and-space}\ \term{then\textvisiblespace} \  \textit{while}\ \term{\textvisiblespace else} \\
	\textit{bool} \to~& \term{false} \gramor \term{\textasciitilde} \ \textit{bool} \gramor \term{true} \gramor \textit{num} \ \term{\textvisiblespace{}==\textvisiblespace{}}\ \textit{num} \\
	\textit{and-space}\to~& \term{\textvisiblespace} \gramor \textit{and-space}\ \term{\&\textvisiblespace}\ \textit{bool}\ \textit{and}\\
 	\textit{num} \to~& \term{L} \gramor \term{n} \gramor \term{(} \ \textit{num}\  \term{+}\  \textit{num}\  \term{)} \\
\end{align*}
\vspace{-2.4em}
\caption{ \ourmethod{}-mined \texttt{while} grammar with 100\% recall.  Nonterminals renamed for readability.}	

\vspace{-1em}
\label{fig:while-grammar-mined}
\end{figure}

\begin{figure}
\vspace{-1em}
	\small
\begin{align*}
\textit{json}\to~& \textit{str} \gramor 
\textit{dict} \ \term{\}} \gramor
\term{false} \gramor 
\term{true} \gramor
\term{[} \ \term{]} \gramor 
\textit{pos-int} \\ &\gramor 
\textit{float-start} \  \regexformat{DIGITS} \gramor
\textit{float-start} \ \textit{pos-int} \gramor
\textit{int} \\ &\gramor 
\term{\{} \ \term{\}}  \gramor 
\term{[} \ \textit{json} \ \textit{list-end} \gramor
\term{null} \gramor 
\regexformat{NAT}\\
\textit{str}\to~& \textit{str-start} \  \term{''} \\
\textit{dict} \to~& \textit{dict-lst} \  \textit{str} \ \term{:} \  \textit{json} 
\hspace{1.3cm}\textit{dict-lst} \to~ \textit{dict} \ \term{,} \gramor \term{\{}\\
\textit{pos-int} \to~& \regexformat{NAT} 
\hspace{3cm}\textit{int}\to  \term{-} \ \textit{pos-int} \gramor \regexformat{NAT}\\
\textit{float-start}\to~&  \textit{int} \ \term{.} \gramor \textit{pos-int} \ \term{.} \\
\textit{list-end}\to~& \term{,} \ \textit{json} \ \textit{list-end} \gramor \term{]}\\
\textit{str-start}\to~& \term{``} \ \textit{chars} \gramor \term{``} \ \textit{pos-int} \gramor \textit{str-start} \ \textit{pos-int}\\
\textit{chars} \to~& \textit{chars} \ \textit{chars} \gramor  \textit{pos-int} \ \textit{chars} \gramor \regexformat{ALNUMS} \\
\regexformat{DIGITS}: \texttt{[0-9}&\texttt{]+} \quad \regexformat{NAT}: \texttt{0|[1-9][0-9]*} \quad  \regexformat{ALNUMS}: \texttt{[a-Z0-9]+}
\end{align*}
\vspace{-1.2em}
\caption{ \ourmethod{}-mined \texttt{json} grammar with maximum F1 Score. Nonterminals renamed for readability. \regexfmt{DIGITS}, \regexfmt{NAT}, and \regexfmt{ALNUMS} are tokens expanded after the  Sec.~\ref{sec:pre-tokenization} pass.}	

\vspace{-1em}
\label{fig:json-grammar}
\end{figure}

The statistics discussed in the prior section show that \ourmethod{}'s mined grammars can closely match the ground-truth grammars in terms of inputs generated and accepted. For RQ5, we consider their human-readable complexity.

Mined grammar readability varies across benchmarks. 
For instance, on the 3 runs where \ourmethod{} achieves 100\% recall for \texttt{while}, the mined grammars look similar to $\grammar_w$ Fig.~\ref{fig:motivating-example}: Fig.~\ref{fig:while-grammar-mined} shows the grammar mined in one of these runs, randomly selected from the three. Fig.~\ref{fig:json-grammar} shows the grammar with maximum F1 score mined by \ourmethod{} on \texttt{json}; it splits some expansions at unusual places (e.g. the use of \textit{float-start}) but is readable after some examination. 

For \texttt{tinyc}, the mined grammars are somewhat over-bubbled: on average they have 56 nonterminals, and 217 rules of average length 1.8. On \texttt{nodejs}, the grammars have on average 40 nonterminals and 276 rules of average length 3.6. 
Because GLADE's grammars are not meant to be human-readable, they are significantly larger: 3505 nonterminals with 4417 rules of average length 1.3 for \texttt{tinyc}; and 2060 nonterminals with 3939 rules of average length 1.2 for \texttt{nodejs}. 


\section{Discussion and Threats to Validity}
\label{sec:discussion}

Our implementation of \ourmethod{} relies on some heuristic elements, which we developed while examining some smaller benchmarks (i.e. \texttt{arith}, \texttt{while}) on a particular set of example strings.  To prevent overfitting on these benchmarks, for evaluation, we used a freshly-generated set of example strings.

The definition of maximal generalization assumes that the language accepted by the oracle is context-free. Thus, we have no formal guarantees on how the algorithm will react to context-sensitive input languages. While our results compared to GLADE are promising, there is no guarantee they will generalize to all benchmarks.

The fact that \ourmethod{}'s \emph{maximum} results consistently beat state-of-the-art (Fig.~\ref{fig:precision-recall-scatter})  suggests a few directions for improvement. If runtime is not a constraint, \ourmethod{} can be parallelized as-is. To choose the winner, first measure precision with respect to the oracle. Then, evaluate the grammars on inputs sampled from the other mined grammars, and choose the one which captures the most of those samples. A less-wasteful way to parallelize would be to conduct some sort of beam search, perhaps using the just-described comparative generalization metric, or to backtrack bad bubbles.

There  remains much room to optimize the order in which bubbles are explored, and pre-tokenization of inputs. We chose two natural metrics for ordering (context similarity and frequency),  but have not exhaustively examined how to combine them. From the difference in performance between the larger benchmarks \texttt{tinyc} (which had simple regex structure) and \texttt{nodejs} (regexes in the training set are more complex), it appears that \ourmethod{} could benefit from running at a higher token level. 
Developing better heuristics for tokenization, or pairing \ourmethod{} with a more complex regex learning algorithm than that described in Section~\ref{sec:pre-tokenization} may yield benefits.




\section{Related Work}
\label{sec:related}
Automatically synthesizing context-free grammars from examples is a long-studied problem in computer science; Lee~\cite{Lee96}, and Stevenson and Cordy~\cite{StevensonSurverySoCP2014} give a survey of some techniques. Gold's theorem~\cite{Gold} states that grammars cannot be learned efficiently from a set of positive examples alone. 
Angluin and Kharitonov~\cite{Angluin95} show that pure black-box approaches face scalability issues on arbitrary CFGs. But, real-world grammars may not be so adversarial. Our heuristics use statistical information to heavily prune the search space. 

The core idea in Solomonoff's~\cite{Solomonoff59} algorithm is to, for each example, find substrings of the example that can be deleted. If a substring can be deleted, Solomonoff proposes to add a recursive repetition rule for the substring. Rather than trying to generalize each example string individually, \ourmethod{} considers all example strings together when producing candidate strings.  Unlike Arvada, Knobe and Knobe~\cite{Knobe76} assume a teacher that can provide new valid strings if the current proposed grammar does not match the target grammar. For each new valid string, their algorithm adds the most general valid 
production of the form $S\to B_1 B_2 \cdots B_n$ to the grammar, where $B_i$ are terminals or existing nonterminal.  It adds new nonterminals by merging nonterminal sequences which have the same left and right contexts in expansions. 
GLADE~\cite{BastaniGladePLDI2017} learns context-free grammars in two phases. First, it learns a regular expression generalizing each input example. Then, it tries to merge subexpressions of these regular expressions in a manner similar to our label merging.  REINAM~\cite{WuReinamFSE2019} uses reinforcement learning to refine a learned CFG, allowing fuzzy matching through a PCFG. It is complementary to our work, as the module that learns a CFG (in their evaluation, GLADE), could be replaced by \ourmethod{}.

$L^*$ and RPNI are two classic algorithms for the learning of \emph{regular} languages. $L^*$ \cite{AngluinLStar1987}  learns regular languages with the stronger assumption of a minimally adequate teacher, which can both (1) act as an oracle for the target language 
, and (2) given a learned regular language, assert whether it is identical to the target language  or  give a counterexample. RPNI~\cite{OncinaRPNI1992} learns regular languages in polynomial time, assuming a set of positive and negative examples. GLADE was found to outperform both these algorithms for program input grammars.
The original $L^*$ paper also describes $L^{cf}$, an algorithm for learning context-free languages in polynomial time, assuming that the set of terminals and non-terminals is known ahead of time. 
This assumption  is not reasonable in most contexts.

Closely related is the field of distributional learning. Clark et. al~\cite{Clark08,Clark10} present polynomial algorithms for learning binary context feature grammars---which capture context-free languages in addition to more complex languages---from strings. The algorithms rely on the representation of words by their \emph{contexts}, an interesting relation to \ourmethod{}'s use of $k$-contexts. Unfortunately, polynomial does not mean fast in practice. We implemented these algorithms in python: even the more efficient one took nearly 5 hours to run on our \texttt{while} benchmark. Work on strong learning~\cite{Clark13} learns grammars with good parse trees---over tokenized inputs. Again, because it uses full context information, it does not scale to large example sets and overgeneralizes on non-substitutable grammars. This highlights the practical importance of $k$-contexts.

Also related is the field of automata learning; learnlib~\cite{Malte15} is a state-of-the-art Java framework implementing several of these algorithms. In particular, it provides an implementation of the TTT~\cite{Malte14} algorithm for learning VPDA. These automata accept a subclass of deterministic context-free languages~\cite{Alur09}. TTT is optimized for situation where the key structure of inputs used to query the oracle can be collected in a prefix-closed set, as in learning from logs of system behavior. This is less well-suited to program inputs with multiple distinct recursive structures. TTT also relies on the stronger assumption of a minimally-adequate teacher, rather than a blackbox oracle.

Another branch of works use grey- or white-box information about the oracle to learn grammars. Lin et al.'s work examines execution traces in order to reconstruct program inputs grammar~\cite{LinA, LinB}.  \textsc{Autogram}~\cite{HoscheleAutogramASE2016} tracks input flows into variables, and uses this dataflow information to learn a well-labeled grammar. Mimid~\cite{GopoinathMimidFSE2020} goes a step further, tracking the control-flow nodes in which input characters are accessed. It directly maps this control-flow structure to the grammar structure, and again can take advantage of function names. The use of this additional oracle information may make the final grammars more robust and speed up the inference process. On the other hand, \ourmethod{}'s blackbox assumption makes it flexible when this information is not readily accessible, or for strangely-structured programs. Our \texttt{tinyc} benchmark was taken directly from Mimid's evaluation, and \ourmethod{} achieved an average F1 score 0.81, compared to Mimid's 0.96. This is impressive given that \ourmethod{} uses the oracle as blackbox.



Section~\ref{sec:deep-learning-eval} discussed the use of deep learning to learn input structures for fuzzing. 
Other techniques do something like grammar mining to increase the effectiveness of fuzzing. Parser-directed fuzzing~\cite{MathisParserDirectedFuzzingPLDI2019} uses direct comparisons to input bytes to automatically figure out tokens of the input structure; it works best on recursive-descent parsers. GRIMOIRE~\cite{BlazytkoGrimoireUSENIX2019} leverages a sort of one-level grammar  by denoting ``nonterminal'' regions of the code as those which can be changed while maintaining a certain kind of branch coverage.

Lastly, the Sequitur compression algorithm resembles the bubbling phase of \ourmethod{}, bubbling sequences that appear with high frequency \cite{NevillManningSequiturJAIR1997}. SEQUIN~\cite{LuhSequinVirology2018} extends Sequitur to mine attribute grammars. Neither algorithm allows for recursive generalization by merging  bubble-induced nonterminals.
 
\section{Conclusion}
\label{sec:conclusion}
We presented \ourmethod{}, a method for learning CFGs from example strings and oracles. We found that \ourmethod{} outperformed GLADE in terms of increased generalization on 11 benchmarks, with a higher F1 score on average on 9 of these benchmarks. These two benchmarks on which \ourmethod{} performs relatively less well are a regular language (for URLs) and a language with more complex regular expressions for tokens. This, along with qualitative analysis of the inputs generated by \ourmethod{} and GLADE, suggests that \ourmethod{} does best in learning recursive structures over tokens, and that a compelling avenue for improvement is a separate token learning step. \ourmethod{} is available as open source at: \url{https://github.com/neil-kulkarni/arvada}.

\section*{Acknowledgements}
Thanks to Rohan Bavishi and all our anonymous reviewers for their invaluable feedback on this paper.  This research is supported in part by gifts from Fujitsu Research of America, and NSF grants CCF-1900968, CCF-1908870,  CNS-1817122.


\bibliographystyle{ieeetr}
\balance
\bibliography{grammars}

\begin{thebibliography}{10}

\bibitem{GopinathF1ArXiV2019}
R.~Gopinath and A.~Zeller, ``{Building Fast Fuzzers},'' {\em CoRR},
  vol.~abs/1911.07707, 2019.

\bibitem{AschermannNautilusNDSS2019}
C.~Aschermann, T.~Frassetto, T.~Holz, P.~Jauernig, A.-R. Sadeghi, and
  D.~Teuchert, ``{Nautilus: Fishing for Deep Bugs with Grammars},'' in {\em
  26th Annual Network and Distributed System Security Symposium}, NDSS '19,
  2019.

\bibitem{Wang19}
J.~Wang, B.~Chen, L.~Wei, and Y.~Liu, ``{Superion: Grammar-Aware Greybox
  Fuzzing},'' in {\em 2019 IEEE/ACM 41st International Conference on Software
  Engineering (ICSE)}, pp.~724--735, 2019.

\bibitem{BastaniGladePLDI2017}
O.~Bastani, R.~Sharma, A.~Aiken, and P.~Liang, ``{Synthesizing Program Input
  Grammars},'' in {\em {Proceedings of the 38th ACM SIGPLAN Conference on
  Programming Language Design and Implementation}}, PLDI 2017, (New York, NY,
  USA), p.~95–110, Association for Computing Machinery, 2017.

\bibitem{WuReinamFSE2019}
Z.~Wu, E.~Johnson, W.~Yang, O.~Bastani, D.~Song, J.~Peng, and T.~Xie,
  ``{REINAM: Reinforcement Learning for Input-Grammar Inference},'' in {\em
  Proceedings of the 2019 27th ACM Joint Meeting on European Software
  Engineering Conference and Symposium on the Foundations of Software
  Engineering}, ESEC/FSE 2019, (New York, NY, USA), p.~488–498, Association
  for Computing Machinery, 2019.

\bibitem{HoscheleAutogramASE2016}
M.~H\"{o}schele and A.~Zeller, ``{Mining Input Grammars from Dynamic Taints},''
  in {\em Proceedings of the 31st IEEE/ACM International Conference on
  Automated Software Engineering}, ASE 2016, (New York, NY, USA), p.~720–725,
  Association for Computing Machinery, 2016.

\bibitem{GopoinathMimidFSE2020}
R.~Gopinath, B.~Mathis, and A.~Zeller, ``{Mining Input Grammars from Dynamic
  Control Flow},'' in {\em Proceedings of the 2019 28th ACM Joint Meeting on
  European Software Engineering Conference and Symposium on the Foundations of
  Software Engineering}, ESEC/FSE 2020, (New York, NY, USA), pp.~1--12,
  Association for Computing Machinery, 2020.

\bibitem{SupplementalTechnicalReport}
N.~Kulkarni, C.~Lemieux, and K.~Sen, ``{Learning Highly Recursive Input
  Grammars: Supplemental Technical Report},'' {\em CoRR}, 2021.

\bibitem{Cummins17}
C.~Cummins, P.~Petoumenos, Z.~Wang, and H.~Leather, ``Synthesizing benchmarks
  for predictive modeling,'' in {\em Proceedings of the 2017 International
  Symposium on Code Generation and Optimization}, CGO '17, p.~86–99, IEEE
  Press, 2017.

\bibitem{ParrANTLRSPE1995}
T.~J. Parr and R.~W. Quong, ``{ANTLR: A Predicated-LL(k) Parser Generator},''
  {\em Software --- Practice \& Experience}, vol.~25, p.~789–810, July 1995.

\bibitem{curl}
D.~Stenberg, ``{cURL: command line tool and library for transferring data with
  URLs}.'' \url{https://curl.se/}, 2018.
\newblock Accessed April 21st, 2021.

\bibitem{URLRFC}
T.~Berners-Lee, L.~Masinter, and M.~McCahill, ``{Uniform Resource Locators
  (URL) }.'' \url{https://tools.ietf.org/html/rfc1738}, 1994.

\bibitem{tinyc}
F.~Bellard, ``{Tiny C Compiler}.'' \url{https://bellard.org/tcc/}, 2018.
\newblock Accessed April 21st, 2021.

\bibitem{nodejs}
O.~Foundation, ``{NodeJS}.'' \url{https://nodejs.org/en/}, 2018.
\newblock Accessed April 21st, 2021.

\bibitem{Zest}
R.~Padhye, C.~Lemieux, K.~Sen, M.~Papadakis, and Y.~Le~Traon, ``Semantic
  fuzzing with zest,'' in {\em Proceedings of the 28th ACM SIGSOFT
  International Symposium on Software Testing and Analysis}, ISSTA 2019, (New
  York, NY, USA), p.~329–340, Association for Computing Machinery, 2019.

\bibitem{LearnFuzz}
P.~Godefroid, H.~Peleg, and R.~Singh, ``{Learn\&Fuzz: Machine Learning for
  Input Fuzzing},'' in {\em Proceedings of the 32nd IEEE/ACM International
  Conference on Automated Software Engineering}, ASE 2017, p.~50–59, IEEE
  Press, 2017.

\bibitem{Cummins19}
C.~Cummins, P.~Petoumenos, A.~Murray, and H.~Leather, ``{Compiler Fuzzing
  through Deep Learning},'' in {\em Proceedings of the 27th ACM SIGSOFT
  International Symposium on Software Testing and Analysis}, ISSTA 2018, (New
  York, NY, USA), p.~95–105, Association for Computing Machinery, 2018.

\bibitem{Lee96}
L.~Lee, ``{Learning of Context-Free Languages: A Survey of the Literature"},''
  tech. rep., Harvard Computer Science Group, 1996.

\bibitem{StevensonSurverySoCP2014}
A.~Stevenson and J.~R. Cordy, ``{A Survey of Grammatical Inference in Software
  Engineering},'' {\em {Science of Computer Programming}}, vol.~96,
  pp.~444--459, 2014.

\bibitem{Gold}
E.~M. Gold, ``{Language Identification in the Limit},'' {\em Information and
  Control}, vol.~10, no.~5, pp.~447--474, 1967.

\bibitem{Angluin95}
D.~Angluin and M.~Kharitonov, ``{When Won't Membership Queries Help?},'' {\em
  J. Comput. Syst. Sci.}, vol.~50, p.~336–355, Apr. 1995.

\bibitem{Solomonoff59}
R.~J. Solomonoff, ``A new method for discovering the grammars of phrase
  structure languages,'' in {\em {Information Processing, Proceedings of the
  1st International Conference on Information Processing}}, pp.~285--289,
  {UNESCO} (Paris), 1959.

\bibitem{Knobe76}
B.~Knobe and K.~Knobe, ``A method for inferring context-free grammars,'' {\em
  Information and Control}, vol.~31, no.~2, pp.~129--146, 1976.

\bibitem{AngluinLStar1987}
D.~Angluin, ``{Learning Regular Sets from Queries and Counterexamples},'' {\em
  Inf. Comput.}, vol.~75, p.~87–106, Nov. 1987.

\bibitem{OncinaRPNI1992}
J.~Oncina and P.~Garcia, ``{Identifying Regular Languages In Polynomial
  Time},'' in {\em Advances in Structural and Syntactic Pattern Recognition},
  vol.~5 of {\em Machine Perception and Artifical Intelligence}, pp.~99--108,
  World Scientific, 1992.

\bibitem{Clark08}
{Alexander Clark and R{{\'e}}mi Eyraud and Amaury Habrard}, ``{A Polynomial
  Algorithm for the Inference of Context Free Languages},'' in {\em Grammatical
  Inference: Algorithms and Applications}, (Berlin, Heidelberg), Springer,
  2008.

\bibitem{Clark10}
A.~Clark, R.~Eyraud, and A.~Habrard, ``{Using Contextual Representations to
  Efficiently Learn Context-Free Languages},'' {\em Journal of Machine Learning
  Research}, vol.~11, no.~92, pp.~2707--2744, 2010.

\bibitem{Clark13}
A.~Clark, ``{Learning Trees from Strings: A Strong Learning Algorithm for some
  Context-Free Grammars},'' {\em Journal of Machine Learning Research},
  vol.~14, no.~75, pp.~3537--3559, 2013.

\bibitem{Malte15}
M.~Isberner, F.~Howar, and B.~Steffen, ``The open-source learnlib,'' in {\em
  Computer Aided Verification} (D.~Kroening and C.~S. P{\u{a}}s{\u{a}}reanu,
  eds.), (Cham), pp.~487--495, Springer International Publishing, 2015.

\bibitem{Malte14}
M.~Isberner, F.~Howar, and B.~Steffen, ``{The TTT Algorithm: A Redundancy-Free
  Approach to Active Automata Learning},'' in {\em Runtime Verification}
  (B.~Bonakdarpour and S.~A. Smolka, eds.), (Cham), Springer International
  Publishing, 2014.

\bibitem{Alur09}
R.~Alur and P.~Madhusudan, ``{Adding Nesting Structure to Words},'' {\em J.
  ACM}, vol.~56, May 2009.

\bibitem{LinA}
Z.~{Lin}, X.~{Zhang}, and D.~{Xu}, ``{Reverse Engineering Input Syntactic
  Structure from Program Execution and Its Applications},'' {\em IEEE
  Transactions on Software Engineering}, vol.~36, no.~5, pp.~688--703, 2010.

\bibitem{LinB}
Z.~Lin and X.~Zhang, ``{Deriving Input Syntactic Structure from Execution},''
  in {\em Proceedings of the 16th ACM SIGSOFT International Symposium on
  Foundations of Software Engineering}, SIGSOFT '08/FSE-16, (New York, NY,
  USA), p.~83–93, Association for Computing Machinery, 2008.

\bibitem{MathisParserDirectedFuzzingPLDI2019}
B.~Mathis, R.~Gopinath, M.~Mera, A.~Kampmann, M.~H\"{o}schele, and A.~Zeller,
  ``{Parser-Directed Fuzzing},'' in {\em Proceedings of the 40th ACM SIGPLAN
  Conference on Programming Language Design and Implementation}, PLDI 2019,
  (New York, NY, USA), p.~548–560, Association for Computing Machinery, 2019.

\bibitem{BlazytkoGrimoireUSENIX2019}
T.~Blazytko, C.~Aschermann, M.~Schl\"{o}gel, A.~Abbasi, S.~Schumilo,
  S.~W\"{o}rner, and T.~Holz, ``{GRIMOIRE: Synthesizing Structure While
  Fuzzing},'' in {\em {Proceedings of the 28th USENIX Conference on Security
  Symposium}}, SEC’19, (USA), p.~1985–2002, USENIX Association, 2019.

\bibitem{NevillManningSequiturJAIR1997}
C.~G. Nevill-Manning and I.~H. Witten, ``{Identifying Hierarchical Structure in
  Sequences: A Linear-Time Algorithm},'' {\em {Journal of Artificial
  Intelligence Research}}, vol.~7, p.~67–82, Sept. 1997.

\bibitem{LuhSequinVirology2018}
R.~Luh, G.~Schramm, M.~Wagner, H.~Janicke, and S.~Schrittwieser, ``{SEQUIN: a
  grammar inference framework for analyzing malicious system behavior},'' {\em
  {Journal of Computer Virology and Hacking Techniques}}, vol.~14, no.~4,
  pp.~291--311, 2018.

\end{thebibliography}
\appendices

\section{Proofs of Existence and Generalization}
\label{sec:concepts}

This appendix provides the proof of the \textbf{\textsc{Existence Theorem}} and the formal statement and proof of the  \textbf{\textsc{Generalization Theorem}} mention in Section~\ref{sec:main-algo}.

\subsection{Definitions}
\label{sec:definitions}

\textbf{Hypothesized Grammar $\mathcal{G_H}$}: Recall that \ourmethod{} keeps a current set of hypothesized trees $\mathcal{T}$. We define the \emph{hypothesized grammar} $\mathcal{G_H}$ of \ourmethod{} to be the grammar induced by $\mathcal{T}$. Just as $\mathcal{T}$ is updated throughout the course of the algorithm, so is $\mathcal{G_H}$.

\textbf{Target Parse Trees}: For the remainder of this section, we refer to the ``hypothesized parse trees'' $\mathcal{T}$ as $P_H$. This notation is made to contrast with the notation $P_O$ for ``target parse trees''. Let $\mathcal{G_O}$ be the target grammar as defined in Section~\ref{sec:technique}. We define the \emph{target parse trees} $P_O$ over the algorithm inputs $S$ as the set of parse trees produced by $\mathcal{G_O}$ when parsing the inputs in $S$.

\textbf{Sufficient Expressivity}: Let $\mathcal{G_O^S}$ be the subgrammar of $\mathcal{G}$ which the examples $S$ can maximally generalize to, as defined in Section~\ref{sec:technique}. We say that $S$ is \emph{sufficiently expressive on $\mathcal{G}$} if $\mathcal{L(G_O^S) = L(G)}$. In other words, a set of sufficiently expressive input examples on $\mathcal{G}$ is a set of examples whose derivations exercise all rules of $\mathcal{G}$.

\textbf{Range of \ourmethod}: We define the \emph{range} of \ourmethod{} as the set $R_A$ of grammars that can be returned by \ourmethod{}.

\textbf{$k$-bubble}: We define a $k$-bubble to be either a single bubble or a double bubble ($k = 1$ and $k = 2$). This concept can be extended to support $k$-bubbles for $k > 3$, but for the purposes of this paper we focus on the case $k \in [1, 2]$.

\subsubsection{Coalesceable Nodes}
\label{sec:coalescable-nodes}

In order for a bubbling operation in the hypothesized parse trees to be accepted, the nonterminal $p$ produced by a single bubble operations must be able to to have its label merged with another existing nonterminal. Similarly, for a double bubble operation, the nonterminals $p_1, p_2$ produced must be able to have their labels merged with each other. We call nodes that are able to have their labels merged \emph{coalesceable} nodes. Therefore, every node created by a bubble operation is \textit{coalesceable}, and \ourmethod{} can only output trees whose interior nodes are coalesceable. The start nonterminal and the terminals at the leaves of the trees are the obvious exceptions to this rule, but we ignore these special cases for the rest of this section, as they will not play into any of the theory.

Let $P$ be a set of parse trees that are only composed of coalesceable nodes. Let the induced grammar on $P$ be $G$. We note that any arbitrary nonterminal $N$ in the grammar $G$ has at least two productions. This is because each nonterminal $N$ in the induced grammar corresponds to the label of some node in $P$. Each node in $P$ is coalesceable and therefore had its label merged with some other node in $P$. This merging of labels creates two rules in the induced grammar.

In Section~\ref{sec:partial-merges}, we introduced the concept of partial merging. While \ourmethod{} only does partial merging in a limited manner for efficiency reasons, assume for the purposes of this proof that we always perform full partial merging, i.e. trying to partially merge all pairs of nonterminals, rather than only partially merging with character nonterminals. When this full partial merging is done, it is not only true that every coalescable node has two productions, but also the reverse implication is true: for any arbitrary grammar $G$, a nonterminal $N$ is \textit{coalescable} if it is the LHS of at least two productions. Without full partial merging, there can be conflicts if two distinct nonterminals have expansions whose right-hand-sides are identical.  Symmetrically, $N$ is \emph{non-coalesceable} if $N$ is the LHS of only one production. 

The rationale behind this definition is intuitive: any valid $k$-bubble that would produce the nonterminal $N$ (e.g. a double bubble that bubbles two of the rules of $N$) will be accepted by the algorithm, since \ourmethod{}'s replacement checks are correct. It's even possible that an incorrect bubble will be accepted, but, as discussed in Section ~\ref{sec:string-sampling}, this unsoundness can be removed with high probability by strengthening our replacement checks.

The above definition of coalesceable nonterminals allows us to extend the concept of coalesceable nodes to arbitrary sets of parse trees $P$. We say $P$ is composed of only coalesceable nodes if the induced grammar on $P$ is composed of only coalescable nonterminals.

\subsubsection{Bubbling Order}
\label{sec:bubbling-order}

\ourmethod{}, as described in Section~\ref{sec:main-algo} performs the following operations in a loop: it gathers the set of all possible bubbles $B$ that can be currently applied to $P_H$, ranks those bubbles, and then for each of those bubbles, tries to see if it leads to a valid merge. If it does, the bubble is accepted and the loop repeats. The behavior of \ourmethod{} is heavily dependent on the order in which bubbles in $B$ are checked.

We say that a bubble $b \in B$ is \emph{prioritized} if it is the first bubble checked in the algorithm loop. Notably, if $b$ leads to a valid merge, the bubble $b$ will be the only bubble in $B$ applied to $P_H$ in that iteration. In Section~\ref{sec:bubble-order} we defined heuristics so that ``better'' bubbles are prioritized. Here, in the Existence Theorem, we show that that for every iteration of the algorithm, there exists a bubble $b$ that if prioritized, will allow \ourmethod{} to fully learn the target grammar. That is, \ourmethod{} will produce as output a grammar $\mathcal{G}$ s.t. $\mathcal{L(G) = L(G_O)}$.

For convenience, we say a sequence of bubbles $(b_i)$ is \emph{prioritized} if on iteration $i$ of \ourmethod{}, bubble $b_i$ is prioritized.

\subsubsection{Strong Graph Isomorphism}
\label{sec:strong-graph-iso}

Let the vertex set of a graph $G$ be denoted $V_G$ and let its edge set be denoted $E_G$. Recall that a graph isomorphism of $G_1$ and $G_2$ is a bijection $f: V_{G1} \rightarrow V_{G2}$ s.t. $\forall (u, v) \in V_{G1} \times V_{G1}, \; (u, v) \in E_{G1} \text{ iff. } (f(u), f(v)) \in E_{G2}$. 

We require a stronger notion of graph isomorphism. Let each node in the graph carry a not-necessarily-unique string ``label''. In a parse tree, a node label would represent the terminal or nonterminal that is stored at that node. We define a \emph{strong graph isomorphism} between $G_1$ and $G_2$ to be a graph isomorphism between $G_1$ and $G_2$, where there is also a one-to-one correspondence between the node labels in $G$ and $H$. 

More rigorously, for any vertex $v$ in a graph $G$, let $v_{label}$ denote the label stored at that vertex. Let the set of node labels in $G_1$ be $L_{G1}$ and the set of node labels in $G_2$ be $L_{G2}$. There is a strong graph isomorphism between $G_1$ and $G_2$ if there is a graph isomorphism $f$ between $G_1$ and $G_2$ and if there exists a bijection $g: L_{G1} \rightarrow L_{G2}$ s.t. $\forall (v, f(v)) \in V_{G1} \times V_{G2}, \; g(v_{label}) = f(v)_{label}$.

\subsection{Proofs}
\label{sec:proofs}

As shown in Section~\ref{sec:coalescable-nodes}, \ourmethod{} can only return a grammar consisting of coalescable nonterminals. Clearly, not all grammars have this property. We show in the next theorem that for any arbitrary target grammar $\mathcal{G_O}$, \ourmethod{} can output a grammar that has a language equivalent to $\mathcal{G_O}$. \\

\textbf{\textsc{Range Lemma:}} For any arbitrary target grammar $\mathcal{G_O}, \; \exists \mathcal{G} \in R \mbox{ such that } \mathcal{L(G)} = \mathcal{L(G_O)}$

\textbf{Proof:} It suffices to show that for any target grammar $\mathcal{G_O} \not\in R_A$, there exists a transformation $\mathcal{T}$ on $\mathcal{G_O}$ s.t. $\mathcal{T(G_O)}$ has the same language as $\mathcal{G_O}$ and $\mathcal{T(G_O)}$ has only coalesceable nodes.

First, we define the transformation $\mathcal{Z}$ as follows. Fix a particular non-coalesceable node $N$ in $\mathcal{G_O}$. Recall that $N$ must have a single production $N \rightarrow \alpha_1 \alpha_2 ... \alpha_m$, where the $(\alpha_i)_{i=1}^m$ is an arbitrary length-$m$ sequence of terminals and nonterminals in $\mathcal{G_O}$. WLOG, assume there is exactly one nonterminal $X$ s.t. $N$ is on the RHS of exactly one production $P_X$ of $X$. The case of multiple productions or multiple such nonterminals that have $N$ on the RHS of a production is handled similarly. Create $\mathcal{Z(G_O)}$ from $\mathcal{G_O}$ by substituting $\alpha_1 \alpha_2 ... \alpha_m$ in place of $N$ in $P_X$. 

Clearly, this transformation is language preserving: that is, $\mathcal{L(Z(G_O)) = L(G_O)}$. Moreover, the non-coalescable nonterminal $N$ is removed from the grammar $\mathcal{G_O}$ after applying the transformation $\mathcal{Z}$.  We define $\mathcal{Z}^k$ as composition of the transformation $Z$ $k$ times. That is, $\mathcal{Z}^k\mathcal{(G) = Z(Z}^{k-1}\mathcal{(G))}$ where $Z^1 := Z$.

Finally, let $\mathcal{G_O}$ have $n$ non-coalesceable nonterminals. We define the transformation $\mathcal{T}$ as $\mathcal{Z}^n$. Since $\mathcal{Z}$ is language-preserving, so is $\mathcal{T}$. And since $\mathcal{Z}$ removes one non-coalesceable node from the grammar, $\mathcal{T}$ removes all non-coalesceable nonterminals from the grammar $\mathcal{G_O}$. $\blacksquare$ \\

\textbf{\textsc{Existence Lemma:}} There exists a sequence $L$ of $k$-bubbles that when prioritized, causes \ourmethod{} to output the target grammar $\mathcal{G_O}$, if $\mathcal{G_O} \in R_A$ and the input examples $S$ are sufficiently expressive on $\mathcal{G_O}$.

\textbf{Proof:} Note that the induced grammar on $P_O$ is the target grammar, since the input examples are sufficiently expressive. Since our algorithm returns the induced grammar on $P_H$, it suffices to show that there exists such a sequence $L$, that when prioritized in \ourmethod{} allows \ourmethod{} to produce hypothesized parse trees $P_H$ s.t. $\forall s \in S, \; P_{Hs} \cong P_{Os}$, where $\cong$ denotes strong graph isomorphism. 

We define the sequence $L$ implicitly by showing that at each point in the algorithm, there exists a bubble that can be made that brings the hypothesized parse trees ``closer'' to strong graph isomorphism with the target parse trees. More rigorously, each bubble of \ourmethod{} produces one or more nodes in $P_H$; let one of these nodes be named $n_H$ in the parse tree $P_{Hs}$ for some $s \in S$. We show that when \ourmethod{} completes, the node $n_H$ produced by the bubble is the preimage under the strong graph isomorphism of some node $n_O$ in $P_{Os}$. We say that this bubble \textit{learned} the node $n_O$ in $P_{Os}$ with the node $n_H$.

Thus, it suffices to show that each iteration of \ourmethod{} learns a node in $P_{Os}$. Since there are a finite number of nodes in $P_{Os}$, after a finite number of iterations of \ourmethod{}, the strong graph isomorphism will be produced. In order to do this, we can show that on each iteration of \ourmethod{}, for any unlearned node $n$ in $P_{Os}$, there exists a bubble $b$ producing nonterminal $p$ that learns $n$. This can be accomplished by induction on the number of iterations of the algorithm, while maintaining the following properties after each iteration:

\begin{itemize}
    \item [\textbf{P1}] The bubble $b$ is valid; that is, the nodes in $b$ are a contiguous subsequence of siblings in the tree $P_{Hs}$ and that the terminal string that induced $b$ is \textit{properly contained}, a term which will be defined later. The bubble $b$ must also lead to a valid merge in order to be accepted
    \item [\textbf{P2}] If any ancestors of $n$ have been learned, let $n_a$, denote the most recent ancestor of $n$ that has been learned. Then, the parent of $p$ is set to be the node that learned $n_a$
    \item [\textbf{P3}] If any children of $n$, $n_c$, have been learned, then $p$ has as children each of the nodes that learned each of the $n_c$
    \item [\textbf{P4}] For any given sibling of $n$, $n_s$, that has been learned, if $n$ is on side $side$ of $n_s$ (where $side$ is either left or right), then $p$ is on side $side$ of the node that learned $n_s$
    \item [\textbf{P5}] If $n$ has a label $l$ that has been learned before, then $p$ is given the label of the node that learned $n$, otherwise $p$ is given an arbitrary fresh label
\end{itemize}

Properties P2-P4 are a formalization of the fact that by the end of the algorithm, node $p$ will be the preimage in a graph isomorphism of $n$, and Property P5 shows that node $p$ will be the preimage in a strong graph isomorphism with node $n$. Property P1 allows us to show that all nodes in the target parse trees will be learned. We propose a mechanism for choosing such a bubble $b$ at each point in the algorithm, and show that all the properties are satisfied by this choice.

First, we define the concept of of a terminal string being \emph{properly contained} in a parse tree. Let $s_s$ be a substring of $s$ from indices $[i, j]$. Note that each nonterminal $n$ in $P_{Hs}$ derives to a substring of $s$ from indices $[i_n, j_n]$. We say that $s_s$ is properly contained if over all $n \in P_{Hs}$, the interval $[i_n, j_n]$ is either contained in $[i, j]$, disjoint from $[i, j]$, or completely contains $[i, j]$. The intuition behind this esoteric definition is that it allows a bubble to be induced by a terminal string, which is described in the following paragraph.

We define the concept of a bubble \emph{induced} by a terminal string in a hypothesized parse tree $P_{Hs}$. Let a terminal string $s_s$ of $s$ from indices $[i, j]$ be properly contained in $P_{Hs}$. Then, there exist a set of nonterminals $N$ that derive to substrings that are contained in $[i, j]$. We take the upper-most nonterminals in $N$ to be the bubble $b_{ss}$ induced by $s_s$. That is, for all $n \in N$ s.t. the parent of $n$ is in $N$, remove $n$ from $N$. Then, $N$ is the bubble induced by $s_s$.

\textbf{Choice of $b$}: Let $n$ be an unlearned nonterminal in $P_{Os}$ s.t. there exists a node $n_l$ that has already been learned that shares the same label as $n$. Let $s_n$ be the string derivable from $n$ in $P_{Os}$. We can let $b$ be the single bubble induced by $n$. Alternatively, we can let $n$ be an unlearned nonterminal in $P_{Os}$ s.t. there is no other node with the same label that has been learned. Since the target grammar $\mathcal{G_O}$ is in the range of \ourmethod{} by assumption, it consists of only coalesceable nonterminals, so there exists another rule with the nonterminal of $n$ as its LHS. This rule must be expressed by the parse trees $P_O$ by the sufficient expressivity condition. Therefore, there exists some other node $n_d$ that is also unlearned with the same label as $n$ in $P_O$. Let $s_n$ and $s_d$ be the strings derivable from $n$ and $n_d$, respectively, and let $b$ be the double bubble induced by $s_n$ and $s_d$, where $s_n$ is applied first if $n$ is lower in the tree $P_{Os}$ than $n_d$, and vice versa. \ourmethod{} doesn't have access to information in the target parse trees, of course, but since it considers all possible bubbles, this choice of bubble will be considered by \ourmethod{}. Here, we make the assumption that it is prioritized as well.

\textbf{Proof by Induction}: In what follows, we show that at each iteration of \ourmethod{} if the bubble $b$ as described above is prioritized, then some previously unlearned node in $P_O$ can be learned by \ourmethod{}. We show that such a node is learned by proving that Properties P1-P5 hold after each iteration of the algorithm, and that the corresponding necessary conditions hold at the start of the algorithm.

\textbf{Inductive Base Case}: We must show that Properties P1-P5 hold at the onset of the algorithm so that our inductive hypothesis is valid. Clearly, Properties P2-P5 trivially hold, since no nodes have been learned. Moreover, since no bubbles have been proposed at the start of the algorithm, the only part of Property P1 that we need to show is that all terminal strings are properly contained. This holds trivially, since the only nonterminal is the start nonterminal which derives to the whole string.

\textbf{Inductive Case (P1)}: We show that the bubble $b$ satisfies Property P1. First, $b$ will certainly lead to a valid merge, and therefore will therefore be accepted. This is because $b$ was constructed based on the strings derivable from nonterminals that are merged in the target grammar.

We assume at the start of the iteration that the terminal string that produced $b$ is properly contained, so we must prove that after application of $b$, the terminal strings of all strings derivable from all of the still-unlearned nodes are also properly contained. A still-unlearned node is either unrelated to $p$, a child of $p$, or a parent of $p$, where $p$ is the node produced by $b$. Let the derivable string of some unlearned-node be $s_u$ and let the derivable strong of $p$ be $s_p$. In the first case, $s_u$ is disjoint from $s_p$; in the second case, $s_u$ encompasses $s_p$; in the third case, $s_p$ encompasses $s_u$, so we have that all necessary strings are still properly contained.

Finally, we must show that $b$ is valid; that is, the nodes in $b$ are siblings in the tree. We know that each node in $b$ has already been learned, by definition. The node $n$ is in the process of being learned, and therefore has not been learned yet, but some ancestors of $n$ may have been learned. Let $n_a$ be the most recent ancestor of $n$ that has been learned. The nodes that constitute $b$ are siblings that share as a parent the node that learned $n_a$, which we call $p_a$. At the time when $p_a$ was learned, any nodes in $b$ that were already learned then were pointed towards $p_a$ by definition of the bubble operation. Nodes in $b$ that were created after $p_a$ was learned were also pointed pointed to $p_a$ as a parent by Property P2.

\textbf{Inductive Case (P2)}: The proof of Property P1 states that all the nodes in $b$ have as a parent $p_a$, where $p_a$ is the node that learned the most recent ancestor $n_a$ of $n$. When bubbling $b$ into $p$, the parent of $p$ is set to be the parent of each of the elements in $b$, which satisfies Property P2.

\textbf{Inductive Case (P3)}: The proof of this case follows from the bubble operation; if a child of $n$ has been learned, then it will by construction appear in the bubble $b$, and each node of the bubble $b$ has its parent set to $p$.

\textbf{Inductive Case (P4)}: The proof of this case follows from the fact that bubbling operations do not change the respective left-right ordering of nodes relative to each other.

\textbf{Inductive Case (P5)}: Following the logic from the proof of Property P1, we know $b$ leads to a valid merge. The correctness of the label name given to the new nonterminal follows from soundness and completeness of the merging operation. Clearly, the merging operation is complete, because for any two nodes that are merged in the target grammar, their derivable strings will replace each other. The merging operation is also sound with high probability if using the  sampling checks described in Section~\ref{sec:string-sampling}.  

\textbf{\textsc{Existence Theorem:}} There exists a sequence of $k$-bubbles, that, when considered by \ourmethod{} in order, enable \ourmethod{} to return a grammar $\mathcal{G}$ s.t. $\mathcal{L(G) = L(\grammar_\oracle)}$, so long as the input examples $S$ are sufficiently expressive on $\mathcal{G_O}$.

\textbf{Proof:} The \textsc{Range Lemma} shows that $\exists \mathcal{G} \in R \mbox{ such that } \mathcal{L(G)} = \mathcal{L(G_O)}$. We take $\mathcal{G_R}$ to be the ``new target grammar''. The construction of $\mathcal{G_R}$ in the \textsc{Range Lemma} should make it clear that if $S$ is sufficiently expressive on $\mathcal{G_O}$, then $S$ is sufficiently expressive on $\mathcal{G_R}$. Applying the \textsc{Existence Lemma} on the target grammar $\mathcal{G_R}$, there exists a sequence $L$ of $k$-bubbles that when considered in order (prioritized), causes \ourmethod{} to output the target grammar $\mathcal{G_R}$, since $\mathcal{G_R} \in R_A$ and the input examples are sufficiently expressive on $\mathcal{G_R}$. Since $\mathcal{G_R}$ is outputted and has a language equivalent to $\mathcal{G_O}$, the claim is proved. $\blacksquare$ \\

\textbf{\textsc{Generalization Theorem:}} $k$-bubbles made on the hypothesized grammar $\mathcal{G_H}$ monotonically increase $\mathcal{L(G_H)}$

\textbf{Proof:} A $k$-bubble is only accepted if it leads to valid merge. We show that this operation monotonically increases the size of the language of the hypothesized grammar. First we define the language of a nonterminal $N$, $\mathcal{L}(N)$, as the set of all strings derivable from the nonterminal $N$. One way to represent the langauge of the whole grammar is to consider the set of all possible strings derivable from the start nonterminal $S$, where the nonterminal $N$ remains unexpanded by terminals; that is, where $N$ is treated as a terminal instead of a nonterminal. We denote this set as $S_N^C$. We have that $\mathcal{L}(G) := S_N^C \times \mathcal{L}(N)$, where the $\times$ denotes performing all possible replacements of strings in $\mathcal{L}(N)$ for the corresponding $N$s in $S_N^C$.

Note that the merge operation merges an arbitrary set of nonterminals $N_1, ..., N_l$ into a new nonterminal $X$, where for any place that a string derivable from $N_i$ could be substituted, a string derivable from some other $N_j$ can also be substituted as well. Recall that for any given nonterminal $N_i$, the the language of the grammar (before the merge) could be defined by $L_{old} := S_{N_i}^C \times \mathcal{L}(N_i)$. However, the language of the grammar after the merge now is the set $L_{new} := S_{N_i}^C \times \cup_j\mathcal{L}(N_j)$, which is a superset of $L$; that is, $L_{new} \supseteq L_{old}$. This gives us that a bubble monotonically increases $\mathcal{L(G_H)}$, as desired. $\blacksquare$

\subsection{Commentary}
\label{sec:commentary}

Our main result is the \textsc{Existence Theorem}, which shows that there exists an ordering of the bubbles which \ourmethod{} follows, it will effectively learn the target grammar $\mathcal{G_O}$, under sufficiently expressive positive examples. In other words, if our algorithm's heuristics prioritize the correct bubbles at each iteration, \ourmethod{} will converge on the correct grammar, given expressive input examples.

Although there are an exponential number of bubbles that can be made in an invocation of \ourmethod, our evaluation in Section~\ref{sec:evaluation} shows that the heuristics implemented in our algorithm allow it to find some ordering of bubbles that allows \ourmethod{} to converge on the correct grammar with a non-negligible probability in a reasonable amount of time.

Finally, the \textsc{Generalization Theorem} shows that even if \ourmethod{} prioritizes an incorrect bubble and therefore is unable to converge on the target grammar, it monotonically generalizes the language of the hypothesized grammar on every iteration.

\end{document}